\documentclass[12pt]{article} 
\usepackage[top=30mm, bottom=41mm, left=30mm, right=30mm]{geometry}

\usepackage{mathtools}
\numberwithin{equation}{section}

\usepackage{indentfirst}
\usepackage{cite} 					

\usepackage{hyperref}
\hypersetup{
	unicode,	
	colorlinks,
	citecolor=[rgb]{0,.3,.5}, 
	linkcolor=[rgb]{0,.3,.5},
	urlcolor=[rgb]{0,.3,.5}, 
	bookmarksopen=true,
	bookmarksopenlevel=\maxdimen,
}

\usepackage{amsmath, amssymb} 
\usepackage{mathrsfs} 			
\usepackage{bm}				

\def\un#1{\underline #1}
\def\vev#1{{\langle #1\rangle}}

\def\on#1#2{{\buildrel{\mkern2.5mu#1\mkern-2.5mu}\over{#2}}}

\def\dt#1{\smash{\on{\hbox{\bf .}}{#1}}\vphantom{#1}}     

\mathcode`\*="702A                  
\def\half{{\textstyle{1\over{\raise.1ex\hbox{$\scriptstyle{2}$}}}}}

\newcommand{\cA}{\mathcal{A}}
\newcommand{\cD}{\mathcal{D}}
\newcommand{\cH}{\mathcal{H}}
\newcommand{\cL}{\mathcal{L}}
\newcommand{\cW}{\mathcal{W}}
\newcommand{\cV}{\mathcal{V}}

\begin{document}
~
\addtolength{\baselineskip}{.5mm}
\thispagestyle{empty}
\vspace{-15mm}
\begin{flushright}
{
{\sc MI-TH-1766}\\
~\\
}
\end{flushright}

\begin{center}
\Large \textbf {
Eleven-Dimensional Supergravity\\
in 4D, $N=1$ Superspace
}
\end{center}

\begin{center}
{
Katrin Becker,
Melanie Becker,
Daniel Butter,
Sunny Guha,
William D. Linch III,
}
\\[3mm]
{\em
George P. and Cynthia Woods
Mitchell Institute for \\
Fundamental Physics and Astronomy, \\
Texas A\&{}M University,\\
College Station, TX 77843, USA
}
\\[3mm]
and
\\[3mm]

Daniel Robbins
\\[3mm]
{\em
Department of Physics, \\
University at Albany, \\
1400 Washington Ave. \\
Albany, NY 12222, USA
}
\end{center}

\vspace{10pt}

\begin{abstract}
We give a formulation of linearized 11D supergravity in 4D, $N=1$
superspace keeping all eleven bosonic coordinates.  The fields are
fluctuations around $\bm M=\mathbf R^{4|4}\times Y$, where $Y$ is a
background Riemannian 7-manifold admitting a $G_2$ structure.  We
embed the 11D fields into superfield representations of the 4D, $N=1$
superconformal algebra.  These consist of the conformal graviton
superfield, seven conformal gravitino
superfields, a tensor hierarchy of superfields describing the 11D
3-form, and a non-abelian Kaluza-Klein vector multiplet gauging the
tensor hierarchy by diffeomorphisms on $Y$.  The quadratic action
consists of the linearization of a superspace volume term and a
Chern-Simons action for the gauged hierarchy coupled to the
supergravity and gravitino superfields, and the full structure is
fixed by superconformal and gauge invariance.  When this action is
projected to components, we recover the full linearized action of 11D
supergravity.
\end{abstract}

\newpage
\tableofcontents

\newpage
\section{Introduction}
\label{S:Intro}

Superstring compactifications on Calabi-Yau three-folds give rise to effective four-dimensional supersymmetric theories which have been extensively studied ever since the seminal work of
Candelas, Horowitz, Strominger and Witten \cite{Candelas:1985en}. As such, they have been an important tool to construct semi-realistic models of particle phenomenology. The massless spectrum is determined by topological properties of the Calabi-Yau manifolds, and if fields are massive, their masses are too large to be observed at least when the compactification scale is high.

More generally, supersymmetric string and M-theory backgrounds are obtained by compactifying on manifolds that admit parallel spinors if fluxes are ignored.  Even though not directly related to semi-realistic models of particle physics, these backgrounds have also been extensively studied. Manifolds with parallel spinors admit a metric with special holonomy, and the choices of possible holonomy groups are enumerated on Berger's list  (cf.\ {\it e.g.}\ \cite{joyce2000compact}).
The holonomy group $SU(3)$, which is relevant for compactifications on Calabi-Yau 3-folds, is only one entry on this list.

In this paper we continue to focus on compactifications of M-theory on {$G_2$}-holonomy manifolds, building off previous work \cite{Becker:2016xgv,Becker:2016rku,Becker:2016edk,Becker:2017njd}.
Our goal is to construct the complete manifestly 4D, $N=1$ supersymmetric space-time action. By complete we mean including all fields (also massive ones) to any order in fluctuations about a background. Supersymmetry is kept explicit by working in superspace.
This is a complex venture not free of problems. The phenomenological ones will not be addressed here (although we have attempted to present our results in a form useful for model building and applications).
These are, of course, very important and M-theory compactifications on {$G_2$}-holonomy manifolds might be a good framework to address vexing physics questions. So for example, it has been argued in ref. \cite{Halverson:2014tya}
that non-perturbative corrections to the space-time action could provide a natural framework to realize de Sitter space in M-(string-)theory. Or it might be
interesting to understand what type of non-abelian gauge groups can be obtained once the {$G_2$}-holonomy manifolds approach a singular limit. The list of interesting phenomenological applications can surely be further expanded and we leave these for future research. In this paper we resolve the theoretical hurdles instead.

Even though the construction of a manifestly supersymmetric space-time action is of interest in its own right, the original motivation came from ref.\ \cite{Becker:2014rea}.  That paper was concerned with the fate of classical solutions of type II or M-theory of the form $\mathbf R^d\times M$, with $M$ a {$G_2$}-holonomy manifold, once perturbative corrections in $\alpha'$ or $1/r$ ($r$ being the radius of $M$) are included.  In particular, is space-time supersymmetry spoiled by these perturbative corrections, or is it possible to perturbatively modify the solution to preserve supersymmetry?  Two approaches were taken.  First, the supersymmetry variation of the internal components of the gravitino was analyzed, and it was shown that the internal metric could be corrected order-by-order to make the variation vanish, provided that a closed 4-form and a closed 5-form were exact at each order.  No $\sigma$-model argument was found to explain why they should be exact.  The second approach to the question was via the D-dimensional effective theory, either 3D, $N=2$ for type II or 4D, $N=1$ for M-theory, including the full Kaluza-Klein  towers of 10D or 11D  fields.  In the language of this effective theory, supersymmetry can only be spoiled by the generation of superpotential or FI terms, but these can be ruled out by combinations of symmetry and holomorphy arguments (the latter relying on the structure of 4D, $N=1$ off-shell superspace).  In fact, a closer analysis of the F- and D-term equations reveals the existence of an exact 4-form and an exact 5-form at each order which were conjectured to correspond to those in the first part.  The expectation was that the two approaches were related by two choices of regularization scheme in the $\sigma$-model.

However, the effective field theory analysis in ref. \cite{Becker:2014rea} was somewhat speculative because the theory had never really been constructed, even in the classical limit ($\alpha'\rightarrow 0$ or $r\rightarrow\infty$).  The arguments in ref. \cite{Becker:2014rea} (as is done there, we will phrase things in 4D, $N=1$ language) involved only vector multiplets and chiral multiplets (and assumed that the 4D 2-forms $C_{mn\, i}$, $m,n$ being 4D indices, $i$ being internal, can be dualized into scalars which sit inside the chiral multiplets), even though there is no consistent action involving these fields alone.  Rather, the assumption was that in the full theory, including also Kaluza-Klein  towers of spin-$\tfrac32$ and spin-2 multiplets, the essential symmetry and holomorphy arguments would remain valid.  There is no reason to believe that this assumption is false, but the absence of an explicit construction of the effective theory, even in the classical limit, is somewhat unsatisfying.  This paper is part of a research program, the goal of which is the construction of the 4D, $N=1$ effective theory that is the classical limit of the theory discussed in ref. \cite{Becker:2014rea}.  Having it in hand would lift the arguments in that paper from speculative to concrete and would open the possibility of studying corrections to the theory, relations to $\sigma$-model regularization schemes, corrections to the geometry of the {$G_2$} moduli space, and other interesting questions.

In refs.\ \cite{Becker:2016xgv, Becker:2016rku, Becker:2016edk} we constructed two important terms of the space-time action in 4D, $N=1$ superspace that, after reducing superfields to components, agreed partially with the space-time action
obtained when 11D  supergravity is compactified on a {$G_2$}-structure manifold.
Even though the superspace and Kaluza-Klein actions were strikingly close (for example, the complete non-linear potential for the metric scalars is reproduced exactly) there were still some differences. These appeared already at the level of field content.
As it turned out, the map between the fields arising from the Kaluza-Klein reduction of  11D  supergravity \cite{Becker:2014uya} and the components of the superfields in refs.\ \cite{Becker:2016xgv, Becker:2016rku, Becker:2016edk} was not one-to-one: 
Some of the latter did not have an M-theory interpretation.
Even then, those fields with an M-theory interpretation were not correctly treated in superspace: When reduced to components some parts of the action in refs.\ \cite{Becker:2016xgv, Becker:2016rku, Becker:2016edk} did not agree with the Kaluza-Klein result.
To characterize these problematic terms, it is easiest to place 4D superfields into representations of the {$G_2$} structure group. Then any kinetic term involving fields in the {\bf 7} representation of {$G_2$}, which could be space-time scalars or gauge fields, did not agree with the Kaluza-Klein result.

In this paper we solve these two problems. Specifically, we show how to accommodate all component fields in the Kaluza-Klein reduction of  11D  supergravity into 4D, $N=1$ superfields without introducing superfluous component fields.
As argued already in ref.\ \cite{Becker:2016edk}, key to this analysis are the gravitino superfields and the new gauge symmetries they imply. As we explain below, this also gives the correct kinetic terms for the fields in the {\bf 7} representation of {$G_2$} from superspace.
In this paper we work to second order in fluctuations about a background given by four-dimensional Minkowski space times a {$G_2$}-holonomy manifold. The non-linear analysis is tractable but will appear in ref. \cite{nonlinear}, since the details would distract from the main points of this paper.

In the next section, we begin with a review of the action as constructed in references \cite{Becker:2016xgv, Becker:2016rku,Becker:2016edk}.
In section \ref{S:Linearized}, we linearize this action around a background of the form $\mathbf R^{4|4}\times Y$ where $Y$ is a Riemannian 7-manifold with fixed $G_2$ structure.
This is then completed by coupling to 4D, $N=1$ conformal supergravity and the previously-missing seven conformal gravitino superfields to quadratic order.
The gauge structure of this action allows for a Wess-Zumino gauge in which only those component fields present in  11D  supergravity remain.
(Our result is summarized in \S{}\ref{S:Assimilation}, to which the reader familiar with such superspace constructions can skip directly.)
In section \ref{S:Components}, we project the linearized action to these components, demonstrating explicitly the matching of the terms and their coefficients to the Kaluza-Klein result.
We recapitulate the salient points of the resulting description of  11D  supergravity in terms of $N=1$ superfields in section \ref{S:CandO} and outline extensions of this result.
Some we are already exploring, such as the full action to linear order in the seven gravitino superfields.
Many other applications have not yet been worked out, but we indicate a few directions for future work we find particularly promising.
Finally, three appendices are included for completeness on technical and quantitative details of $G_2$ geometry (\S{}\ref{S:G2}), linearized 4D, $N=1$ supergravity (\S{}\ref{S:OMSG}), and 4D, $N=1$ conformal gravitino superfields (\S{}\ref{S:4DGravitino}).
(The analysis in appendix \ref{S:4DGravitino} was carried out for general gravitino representations and has applications to similar constructions in dimensions other than eleven.)

\section{Review and Overview}
\label{S:Review}

Our goal is to embed the components of  11D  supergravity into 4D, $N=1$ superfields and to construct the complete two-derivative  11D  action in terms of them.
In this section, we review those parts of the construction already worked out in refs.  \cite{Becker:2016xgv, Becker:2016rku,Becker:2016edk}.
In section \ref{S:11Decomposition}, we recall how to reduce the components of  11D  supergravity to 4+7 dimensions. These are embedded into $N=1$ superfields. The superfields accommodating the  11D  3-form give rise to a tensor hierarchy. In section  \ref{S:NATH} we recall that the
structure of this tensor hierarchy uniquely fixes the superspace Chern-Simons action. 
Similarly, the K\"ahler action is strongly constrained, as we discuss in section \ref{S:Kahler}.
The resulting component action reproduces many terms of the 11D action.
In section \ref{S:Faithful} we describe the remaining mismatch and how it will be resolved.

\subsection{Decomposition of Eleven-Dimensional Supergravity}
\label{S:11Decomposition}

The components of 11D supergravity consist of the frame field $e_{\bm m}{}^{\bm a}$, the gravitino field $\psi_{\bm m}^{\bm \alpha}$, and the 3-form $C_{\bm{mnp}}$.
Here ${\bm m, \bm n\dots} = {0,\dots,10}$ are coordinate indices, ${\bm a, \bm b \dots} = {0,\dots,10}$ are tangent indices, and ${\bm \alpha, \bm  \beta \dots}={1,\dots, 32}$ are Majorana spinor indices.
The two-derivative 11D supergravity action is given by \cite{Cremmer:1978km}
\begin{align}
\label{E:CJS}
\kappa^ 2 S_{11} &= \frac12\int d^{11}x \, e \left(
		R 
		- \frac{1}{48} (F_{\bm{mnpq}})^2
	\right)
-\frac1{12} \int C \wedge F \wedge F
\\ &	
	+\frac12\int d^{11}x \, e \left(
	-i \bar\psi_{\bm m} \gamma^{\bm {mnp}} {{\mathcal D}}_{\bm n} \psi_{\bm p}
	+\frac 1{192} \bar\psi_{\bm r} \gamma^{\bm r} \gamma^{\bm {mnpq}} \gamma^{\bm s}  \psi_{\bm s} F_{\bm{mnpq}}
	\right)
+\dots
\nonumber
\end{align}
where we suppress $\psi^4$ terms for simplicity.
Here $R$ and $\mathcal D$ are the Ricci scalar and covariant derivative constructed from the frame and spin connection, and $F=dC$ is the 4-form field strength. 

Locally we treat the 11D space-time as a direct product $X \times Y$
with $X$ and $Y$ describing 4- and 7-dimensional manifolds respectively.
This involves decomposing
coordinate indices as $\bm m \to m,i$ with $m,n,\cdots$
denoting $GL(4)$ indices and $i,j,\cdots$ denoting $GL(7)$ indices;
tangent space indices as $\bm a \to a,\underline{i}$ with $a,b,\dots = 0,\dots, 3$
denoting $SO(1,3)$ and $\underline{i}, \underline{j}, \cdots = 1, \cdots, 7$ denoting $SO(7)$ indices;
and spinor indices as $\bm \alpha \to (\alpha\otimes I, \dt \alpha\otimes I)$, with
$\alpha$ and $\dt \alpha$ denoting chiral and antichiral $Spin(1,3)\cong SL(2;\mathbf C)$
indices and $I$ denoting a $Spin(7)$ index.

If all supersymmetries are kept on equal footing, it is natural to augment the
$Spin(7)$ symmetry (which is an R-symmetry from the 4D perspective) to $SU(8)$
\cite{deWit:1986mz}. This matches the full R-symmetry group of 4D, $ N=8$
supergravity, both ungauged \cite{Cremmer:1979up} and gauged
\cite{deWit:2007kvg}, and permits the scalar fields of the theory to be
interpreted as parameterizing an $E_{7(7)} / SU(8)$ coset space.
Then the eight gravitini
and 56 spin-$\tfrac12$ fermions are grouped into irreducible representations of $SU(8)$.
If instead the goal is to maintain manifest 4D, $N=1$ supersymmetry, one of the eight gravitini
$\psi_m^{\alpha I}$ must be separated from the rest. 
This can be
achieved by taking
\begin{align}
\psi_{m}^{\alpha I} = \psi_{m}^{\alpha} \eta^I
	+ i \psi_{m i}^{\alpha} (\Gamma^{i})^{I J} \bar \eta_J
{~~},
\end{align}
where $\eta^I$ is a fixed complex spinor that selects out the preferred $N=1$ supersymmetry
and $\Gamma^i = \Gamma^{\underline{j}} e_{\underline{j}}{}^i$ in terms of the $Spin(7)$ gamma matrix $\Gamma^{\underline{i}}$.
The additional seven gravitini $\psi_{m i}^{\alpha}$ now carry a $GL(7)$ index.
A similar decomposition applies to the other 56 fermions, leading to spinors
$\chi_{ij}^\alpha$ and $\chi_{ijk}^\alpha$, which are totally antisymmetric in their
$GL(7)$ indices. Dealing with the internal components $g_{i j}$ of the metric
rather than with the internal vielbein, all fields can thus be chosen to carry $GL(7)$
indices rather than $SO(7)$ or $Spin(7)$ indices.
We summarize this field content in table \ref{T:BoseSpectrum}.
\begin{table}[t]
{
\hspace{-6mm}
\begin{tabular}{|c|c|c|c|c|c|c|c|}
\hline
	 & metric & gravitino & 3-forms & 2-forms & (axial-)vectors & (pseudo-)scalars & spinors \cr
\hline
$g_{\bm{mn}}$ & $g_{mn}$&|&|&|& $g_{mi}$ & $g_{ij}$ &| \cr
1 & 1& 0& 0 & 0 & 7 & 28 & 0 \cr
$C_{\bm{mnp}}$ &|&|& $C_{mnp}$ & $C_{mni}$ & $(C_{mij})$ &$(C_{ijk})$&|  \cr
1 & 0& 0& 1 & 7 & 21 & 35 & 0\cr
$\psi_{\bm{m}}^{\bm{\alpha}}$ &|& $\psi_m^\alpha$, $\psi^{\alpha}_{m i}$ &|&|&|&|& $\chi^\alpha_{ij}$, $\chi^\alpha_{ijk}$ \cr
1 & 0& $1+7$ & 0 & 0 & 0 & 0 & $21+ 35$\cr
\hline
\end{tabular}
} 
\begin{caption}{Component spectrum for the $11 \to 4+7$ split.}
\footnotesize
The reduction of 11D supergravity fields gives rise to a mixture of forms of various degrees. After dualizing axial-vectors and 2-forms, we find a total of $7+21 = 28$ vectors, $28+7 = 35$ scalars, and 35 pseudo-scalars of $N=8$ supergravity \cite{Cremmer:1979up}. The 3-form is non-dynamical on-shell. Axial-vectors and pseudo-scalars are marked with parentheses.
\label{T:BoseSpectrum}
\end{caption}
\end{table}

To count degrees of freedom it is convenient to dualize the seven space-time 2-forms $C_{mn i}$ into seven scalars.
When combined with the 28 metric scalars $g_{ij}$, they can be arranged into a 3-form $\varphi_{ijk}$.
Invertibility of the metric requires this 3-form to be non-degenerate in a suitable sense (reviewed in appendix \ref{S:G2}).
When pulled back to $Y$, this condition defines a $G_2$ structure \cite{Hitchin:2000jd, Hitchin2001}. Together with the 35 pseudo-scalars $C_{ijk}$, these gravitational scalars can embedded into superspace as the lowest components of a chiral pseudo-scalar superfield $\Phi_{ijk}$.

Note, however, that we will not be dualizing the seven space-time 2-forms into scalars, because this procedure obscures the gauge structure of the tensor hierarchy.
Nevertheless, we will still embed the 28 metric scalars and 35 pseudo-scalars into a chiral superfield $\Phi_{ijk}$. Consequently there are $35-28=7$ additional scalars not accounted for
in the list of fields obtained by dimensional reduction. We will later see that these additional seven scalars are pure gauge degrees of freedom.
Similarly, we embed the remaining components of the 11D 3-form into an abelian tensor hierarchy of superfields \cite{Becker:2016xgv}.
This is a chain complex of superforms constructed from the de Rham complex on $Y$ tensored with the super-de Rham complex on $\mathbf R^{4|4}$.
The forms in this complex are charged under diffeomorphisms on $Y$ and so couple to the non-abelian Kaluza-Klein vector field $\mathcal A_a^i = - e_a{}^i$.
Any such complex of superforms, also known as a non-abelian tensor hierarchy, has a Chern-Simons-like invariant $S_{CS}$ \cite{Becker:2016rku, Becker:2017njd}.
We now turn to the details of this embedding.

\subsection{3-form Hierarchy and Chern-Simons Action in Superspace}
\label{S:NATH}

Four-dimensional $p$-forms are embedded into superfields as summarized in table \ref{T:superforms} \cite{Gates:1980ay,Gates:1983nr}.
\begin{table}[t]
{\renewcommand{\arraystretch}{1.3} 
\hspace{-9mm}
\begin{tabular}{|c|c|c|c|c|c|}
\hline
$p$ & lowest component & constraints & prepotential & top component  \cr
\hline
0 & $F_\alpha = D_\alpha F$
	& $\bar D^2  D F=0$
	& $F = \tfrac1{2i}(\Phi - \bar \Phi)$
	& $F_{a} = \sigma_a [D, \bar D] F $ \cr
1 & $F_{\alpha a} = (\sigma_a)_{\alpha \dt \alpha}  \bar W{}^{\dt \alpha}$  
	& $D \bar W=0$ \& $\bar D \bar W=D W$
	& $W = -\tfrac14\bar D^2 D V$
	& $F_{ab} = D\sigma_{ab} W +\mathrm{c.c.}$ \cr
2 & $F_{\alpha\dt\alpha a} = (\sigma_a)_{\alpha \dt \alpha} H $
	&
	$D^2 H=0$
	& $H =\tfrac1{2i}( D \Sigma-\bar D \bar \Sigma)$
	& $F_{abc} = \epsilon_{abcd} \sigma^d
		[D, \bar D]H$ \cr
3 & $F_{\alpha\beta ab} = (\sigma_{ab})_{\alpha \beta} G $
	& $\bar D G=0$
	& $G = -\tfrac14 \bar D^2 X$
	& $F_{abcd} = i \epsilon_{abcd} D^2 G +\mathrm{c.c.} $ \cr
\hline
\end{tabular}
} 
\begin{caption}{Conventional embedding of $p$-forms in closed superforms.}
\footnotesize
The component $p$-forms are embedded into closed super $(p+1)$-form field strengths $F$ as originally shown in \cite{Gates:1980ay}. Each field strength can be written in terms of an invariant scalar or spinor superfield. These satisfy constraints that can be solved in terms of prepotentials. Overall numerical constants have been neglected in the first and last columns.
\label{T:superforms}
\end{caption}
\end{table}
As a guiding principle it is 
useful to note that the gauge transformation of a $p$-form is formally identical to the field strength of a $(p-1)$-form and similarly for their Bianchi identities.
In terms of super-$p$-forms\footnote{Here $A_1,A_2,\dots $ are 4D, $N=1$ superspace indices. } $F_{A_1\dots A_p}$, the lowest-dimension non-vanishing component is indicated in the first column.
Analyzing the Bianchi identities, the higher components are found in terms of superspace derivatives of the lowest component \cite{Gates:1980ay,Gates:1983nr} (see also app. A of ref. \cite{Becker:2017njd}).

This description of $p$-forms in 4D, $N=1$ superspace can be extended to accommodate both the dependence on the additional seven coordinates \cite{Becker:2016xgv} and the minimal coupling to the non-abelian gauge field of the Kaluza-Klein vector \cite{Becker:2016rku}. First, the non-abelian connection $\cA^i$ is added to the $N=1$ superspace derivative,
\begin{align}
\label{E:CovDred}
\cD_C &= D_C - \mathscr L_{\cA_C}{~},
\end{align}
where $\mathscr L$ denotes the Lie derivative on $Y$. This can be separated into the de Rham differential $\partial$ on $Y$ and the contraction operator $\iota$ using Cartan's formula,
$\mathscr L_\mathcal V = \partial \iota_\mathcal V + \iota_\mathcal V \partial$.
The field strength $\cW_\alpha{}^i$ of the non-abelian connection is defined by
\begin{align}
[\mathcal D_a, \bar {\mathcal D}_{\dt \alpha}] = - (\sigma_a)_{\alpha \dt \alpha} \mathscr L_{\mathcal W^{\alpha}}.
\end{align}
This definition implies the conditions \cite{Gates:1983nr, Wess:1992cp, Buchbinder:1998qv}
\begin{align}
\bar\cD_{\dt \alpha} \cW_\alpha{}^i = 0
~~~\textrm{and}~~~
\cD^{\alpha} \cW_{\alpha}{}^i = \bar\cD_{\dt\alpha} \bar\cW^{\dt\alpha}{}^i~.
\end{align}
The invariant field strengths of the $p$-form hierarchy can be written
in terms of prepotentials as
\begin{subequations}
\label{E:NATHFS}
\begin{align}
E &= \partial \Phi
\\
\label{E:FSF}
F &= \tfrac1{2i}\left( \Phi - \bar \Phi\right) - \partial V
\\
W_\alpha &= -\tfrac14 \bar \cD^2 \cD_\alpha V
	+\partial \Sigma_\alpha
	+ \iota_{\mathcal W_\alpha} \Phi
\\
H &= \tfrac1{2i}\left(\cD^\alpha \Sigma_\alpha - \bar \cD_{\dt \alpha} \bar \Sigma^{\dt \alpha} \right)
	-\partial X
	-\omega(\mathcal W_\alpha, V)
\\
\label{E:G}
G&= -\tfrac14 \bar \cD^2 X
	+ \iota_{\mathcal W^\alpha} \Sigma_\alpha
{~~}.
\end{align}
\end{subequations}
Here, $\Phi$ and $\Sigma_\alpha$ are chiral superfields while
$V$ and $X$ are real unconstrained superfields.
All fields are differential forms on $Y$.
The composite superfield $\omega$ is the Chern-Simons superfield; for any $p$-form scalar superfield $v$,
\begin{align}\label{E:CSsuperform}
\omega(\cW_\alpha, v) &:=
\iota_{\cW^\alpha} {\cD}_\alpha v
	+ \iota_{\bar \cW_{\dt \alpha}} \bar {\cD}^{\dt \alpha} v
	+\tfrac12 \left(  \iota_{{\cD}^\alpha\cW_\alpha} v
		+ \iota_{\bar {\cD}_{\dt \alpha} \bar \cW^{\dt \alpha}} v
	\right)~.
\end{align}
Its name derives from the fact that 
$-\tfrac{1}{4} \bar \cD^2 \omega(\cW_\alpha, v) = \iota_{\cW^\alpha} \chi_\alpha$, where $\chi_\alpha = -\tfrac{1}{4} \bar \cD^2 \cD_\alpha v$ is the field strength superfield of $v$, is a product of field strengths. 
(This is then the superspace analog of $d \omega \sim \mathcal F\wedge F$.)

The prepotential superfields transform covariantly under both non-abelian internal diffeomorphisms as well as the abelian tensor hierarchy gauge transformations
\begin{subequations}
\label{E:NATHtransformations}
\begin{align}
\delta \Phi &=
	\mathscr L_\tau \Phi
	+ \partial \Lambda
\\
\delta V &=
	\mathscr L_\tau V
	+ \tfrac1{2i}\left(\Lambda - \bar \Lambda \right)
	- \partial U
\\
\delta \Sigma_\alpha &=
	\mathscr L_\tau \Sigma_\alpha
	-\tfrac14 \bar {\cD}^2 {\cD}_\alpha U
	+ \partial \Upsilon_\alpha
	+ \iota_{\mathcal W_\alpha} \Lambda
\\
\delta X &=
	\mathscr L_\tau X
	+\tfrac1{2i}\left({\cD}^\alpha \Upsilon_\alpha - \bar {\cD}_{\dt \alpha} \bar \Upsilon^{\dt \alpha} \right)
	- \omega(\mathcal W_\alpha, U)~.
\end{align}
\end{subequations}
The abelian part of the gauge transformation is parameterized by the superfields $\Lambda_{ij}$ (chiral), $U_i$ (real), and $\Upsilon_\alpha$ (chiral) encoding the components of an eleven-dimensional super-2-form. The non-abelian parameter $\tau^i$ is a real superfield describing internal diffeomorphisms. The field strengths \eqref{E:NATHFS} are invariant under the abelian transformations but transform as $p$-forms under internal diffeomorphisms.

Having been given explicitly in terms of the prepotential superfields, the field strengths identically satisfy the Bianchi identities
\begin{subequations}
\label{E:BI}
\begin{align}
0 &= - \partial E
	\\
\tfrac1{2i}\left(E - \bar E\right) &=  \partial F
	\\
-\tfrac14 \bar {\cD}^2 {\cD}_\alpha F  &=
	-\partial W_\alpha
	- \iota_{\mathcal W_\alpha} E
	\\
\label{E:NATHBI1W}
\tfrac1{2i}\left({\cD}^\alpha W_\alpha - \bar {\cD}_{\dt \alpha} \bar W^{\dt \alpha} \right)&=
	\partial H
	+ \omega(\mathcal W, F)
	\\
-\tfrac14 \bar {\cD}^2 H &=
	-\partial G
	- \iota_{\mathcal W^\alpha} W_\alpha
	\\
\bar {\cD}_{\dt \alpha} G &= 0~.
\end{align}
\end{subequations}
Just as in 4D, $N=1$, these identities are the superspace analogs of $dF = 0$ \cite{Becker:2017njd}.

The Chern-Simons action $\int C \wedge F \wedge F$ can be embedded in superspace by first constructing the superspace analogue of the closed 8-form $F\wedge F$. When rewritten in 4+7 dimensions, this form generates a hierarchy of 4D $p$-forms with $8-p$ internal indices for $p=1$ through $p=4$. These four $p$-forms are embedded into $N=1$ superfields as \cite{Becker:2017njd}
\begin{subequations}
\begin{align}
\label{E:NACS2}
\mathbb F &= \tfrac{1}{2} (E + \bar E) F
\\
\mathbb W_\alpha&= E W_\alpha -\tfrac i4 \bar {\cD}^2 (  F {\cD}_\alpha F)
\\
\mathbb H &= \tfrac{1}{2} (E +\bar E) H + \omega(W, F)
	- i {\cD}^\alpha F\iota_{\mathcal W_\alpha}F
	+ i \bar {\cD}_{\dt \alpha} F \iota_{\bar{\mathcal W}^{\dt \alpha} }F
\\
\mathbb G&= E G + \tfrac12 W^\alpha W_\alpha - \tfrac i4 \bar {\cD}^2 (  F H)
\end{align}
\end{subequations}
with wedge products suppressed. Here, the Chern-Simons superform $\omega(W, F)$ is constructed analogously to
\eqref{E:CSsuperform} but with contraction replaced by wedge product: 
\begin{align}
\omega(W_\alpha, v) &:=
	W^\alpha \wedge {\cD}_\alpha v
	+ \bar \cW_{\dt \alpha} \wedge \bar {\cD}^{\dt \alpha} v
	+\tfrac{1}{2} {\cD}^\alpha W_\alpha \wedge v
	+ \tfrac{1}{2} {\bar {\cD}_{\dt \alpha} \bar \cW^{\dt \alpha}} \wedge v~.
\end{align}
These satisfy the descent relations (\ref{E:BI}) expressing the fact that the superform $F\wedge F$ is closed in the extended super-de Rham complex provided $F$ is \cite{Becker:2017njd}.
Because of these closure relations, the action
\begin{align}
S_{CS} &= {1\over\kappa^ 2}\int d^4 x \int_Y L_{CS}{~}, \nonumber \\
\label{E:NACS}
L_{CS} &=
	-\frac i{12} \int d^2 \theta  \,  \left(  \Phi \mathbb G + \Sigma^\alpha \mathbb W_\alpha
	\right)
	- \frac 1{12}\int d^4 \theta  \,  \left( V \mathbb H - X \mathbb F
	\right)
	+\mathrm{h.c.} 	
\end{align}
is invariant under the non-abelian tensor hierarchy transformations (\ref{E:NATHtransformations}).

\subsection{Symmetries of the Superspace Embedding}
\label{S:Symmetries}

Eleven-dimensional supergravity is invariant under the global ``trombone'' scaling symmetry \cite{Cremmer:1997xj}. This means one may assign engineering dimension to each 11D component field so that the 11D Planck constant $\kappa^2$ appears as an overall factor, as we have already chosen in eqn. \eqref{E:CJS}. Of course, the 4D Chern-Simons action must retain this feature; but, moreover, it possesses a global 4D, $N=1$ superconformal symmetry, which is due to its $p$-form origin. In table \ref{T:Weights}, we give the global conformal $(\Delta)$ and chiral $U(1)_R$ ($w$) weights as well as the engineering dimension ($d$) of the various fields and operators.
They are useful when studying superconformal interactions beyond the linearized approximation and in constructing models with this field content.

\begin{table}[t]
\begin{align*}
{\renewcommand{\arraystretch}{1.3} 
\begin{array}{|c|ccccccc|ccccc|}
\hline
	&\kappa^2& \int d^7y &\int d^4x & \int d^4 \theta & \int d^2 \theta& \cD_\alpha & \partial_i
	& G(X) & H(\Sigma) & W_\alpha(V) &  F(\Phi) & \mathcal W(\mathcal V) \\
\hline
\Delta &0& 0& -4& 2 & 1& \tfrac{1}{2} & 0 & 3 (2) & 2 (\tfrac32) & \tfrac32(0) & 0(0) & \tfrac32(0)\\
w &0& 0&0 & 0 &-2 & -1 & 0 & 2(0) 	&  0 (1) 	& 	1(0)	& 0(0)& 	1(0)\\
d  &-9& -7 &-4& 2& 1 & \tfrac{1}{2} & 1 & 0(-1)	&	0(-\tfrac12)& \tfrac12(-1) & 0(0)&\tfrac12(-1)\\
\hline
\end{array}
}
\end{align*}
\begin{caption}{Superconformal weights ($\Delta$ and $w$) and engineering dimension ($d$) of various objects}
\label{T:Weights}
\end{caption}
\end{table}

\subsection{K\"ahler Action}
\label{S:Kahler}
\label{S:KahlerReview}

The Chern-Simons action \eqref{E:NACS} contains at the component level the 11D Chern-Simons action as well as kinetic terms for the vector fields. But to recover the kinetic terms for the scalars and 2-forms, an appropriate K\"ahler term must be added. Keeping in mind the superconformal weights and engineering dimensions of table \ref{T:Weights}, the K\"ahler term can be specified up to an undetermined function $\mathcal H$ as
\begin{align}
\label{E:Kahler}
S_K = -\frac3{\kappa^2} \int d^4 x \int d^7y \int d^4\theta  \sqrt{g(F)}  \,
	(\bar GG)^{1/3} \,\mathcal H~.
\end{align}
Here $g(F) = \mathrm{det}(g_{ij}(F))$ is the determinant of the Riemannian metric on $Y$ obtained from the $G_2$-structure 3-form by replacing $\varphi_{ijk}$ with the superfield $F_{ijk}$ for which it is the background value. This Hitchin-like functional \cite{Hitchin:2000jd, Hitchin2001} serves here as a measure term necessary for covariance under internal diffeomorphisms. The factor of $(G \bar G)^{1/3}$ is chosen to provide the appropriate superconformal weight. The remaining function $\cH$ must be weight-less and transform as a scalar under internal diffeomorphisms. For a two-derivative action, it may depend only on $F_{ijk}$, $G$, and $H_{i}$ and, in order to be a weight-less scalar, only in a specific combination:
\begin{align}
\cH = \cH(x)
~~~\textrm{with}~~~
x := (\bar G G)^{-2/3} g^{ij} H_i H_j~.
\end{align}
The uniqueness of 11D supergravity means that the function $\cH(x)$ must be fixed by extended supersymmetry. In a forthcoming publication \cite{nonlinear}, we confirm this, give its explicit form, and explain its origin from 11D. 
As we will be working to quadratic order in this paper, we will need only the two lowest-order terms in the $x$-expansion.
Reference \cite{Becker:2016edk} showed that matching certain kinetic terms in 11D supergravity as well as the scalar potential required that $\cH(0) = 1$. We define the first correction
\begin{align}
\label{E:c}
c := \mathcal H'(0)
{~}.
\end{align}
Requiring invariance under linearized extended supersymmetry (\S{}\ref{S:ExtendedSUSY}) will fix $c=-\tfrac14$.

A few comments are in order. First, the chiral superfield $G$ is closely related to the conformal compensator of 4D, $N=1$ supergravity. Its conformal weight and its role in the action suggest its identification  as $\Phi_0^3$ where $\Phi_0 = e^{\sigma}$ is the chiral superfield compensator in old minimal supergravity. The fact that $G$ is built from a real prepotential $X$ rather than a complex one means that it is a constrained chiral multiplet: One of its auxiliary scalar fields is replaced by the dual of the 4-form field strength $F_{mnpq}$ of eleven-dimensional supergravity.
We review the construction of old minimal supergravity and this modification in appendix \ref{S:OMSG}.

\subsection{Faithful Embedding and Compensating Superfields}
\label{S:Faithful}

\begin{table}[t]
{\renewcommand{\arraystretch}{1.2} 
\begin{center}
\hspace{-5mm}
\begin{tabular}{|c|c|c|c|c|c|c|c|}
\hline
	& 3-forms 	& 2-forms & vectors		& scalars 	& spinors & auxiliaries \\
\hline
$X$	& $C_{mnp}$	& | 		& | 			& $G$				& $\zeta^\alpha$ & $d_X$ \\
$\Sigma^\alpha_i$ & |& $C_{mni}$& |		 	& $H_{i}$			& $\zeta^\alpha_{i}$ & | \\
$V_{i j}$ & |		& |		& $C_{m i j}$	& |					& $\chi^\alpha_{i j}$ & $d_{i j}$ \\
$\Phi_{ijk}$& |	& |		& |			& $C_{i j k}$, $F_{i j k}$ & $\chi^\alpha_{ijk}$ & $f_{i j k}$\\
$ \cV^i$ & |		& |		& $g_{m i}$	& |					& $\zeta^{\alpha i}$ & $\bm{d}^i$ \\
\hline
\end{tabular}
\end{center}
} 
\begin{caption}{Component spectrum of Chern-Simons prepotentials}
\label{T:PotentialSpectrum}
\footnotesize
We present the component fields contributed by the gauged Chern-Simons hierarchy. 
As explained in the text, the bosons $G$, $H_i$, and seven of the $F_{ijk}$ can all be removed by a choice of Wess-Zumino gauge.
Similarly, the fermions $\zeta$ can be gauged away or vanish on-shell (by the auxiliary field $\rho$ of the gravitino multiplet).
\end{caption}
\end{table}

We have embedded the components of eleven-dimensional supergravity up through spin-1 into 4D, $N=1$ superspace, but in doing so we have been forced to introduce additional components not present in the 11D spectrum of table \ref{T:BoseSpectrum}. We list in table \ref{T:PotentialSpectrum} the component fields of the various superfields of the non-abelian tensor hierarchy. Compared to the 11D spectrum, there are 16 additional scalar fields: two from the lowest component of the chiral superfield $G$, seven from $H_i$, and another seven from $F_{ijk}$. The latter arise because, as mentioned at the end of section \ref{S:11Decomposition}, $F_{ijk}$ should encode only the 28 degrees of freedom of the internal metric. (Using the background value of $F_{ijk}$ as the $G_2$-structure 3-form, the scalars in $F_{ijk}$ may be decomposed into $G_2$ representations as $\bm {35} = \bm {27} \oplus \bm {7} \oplus \bm {1}$. The troublesome scalars are contained in the $\bm{7}$.)
There is a similar surfeit of fermions. Aside from the expected 56 fermions $\chi_{ij}$ and $\chi_{ijk}$, there are 15 more: $\zeta$, $\zeta_i$, and $\zeta^i$. In addition to this overcounting of the spin $\leq 1$ degrees of freedom, the $N=1$ description does not yet encompass the spin-$\tfrac32$ and spin-2 degrees of freedom.

It was proposed in \cite{Becker:2016edk} to introduce the graviton by minimally coupling to $N=1$ conformal supergravity. This can be done either covariantly---by introducing an appropriate measure $E$ and defining a curved space covariant derivative $\nabla_A$ to replace $\cD_A$---or by explicitly coupling to the gravitational prepotential $H^{a}$ \cite{Gates:1983nr,Buchbinder:1998qv}.
As we review in appendix \ref{S:OMSG}, this is a real, Lorentz 4-vector-valued superfield analogous to the Yang-Mills prepotential. Using the Pauli matrices $(\sigma_a)_{\alpha \dt \alpha}$, we can write any real vector as a Hermitian matrix $H_{\alpha \dt \alpha}$.\footnote{In this work, we will freely switch back and forth between real 4-vectors and Hermitian matrices using $H_{\alpha \dt \alpha}=\sigma^a_{\alpha \dt \alpha} H_a$ and $H_a=-\tfrac{1}{2}\bar{\sigma}_a^{\dt \alpha \alpha}H_{\alpha \dt \alpha}$.
In such conversions, contractions give factors of $-2$: $H_{\alpha \dt \alpha} H^{\alpha \dt \alpha} =  -2 H_a H^a$. }
Under linearized local superconformal transformations, this superfield transforms as
\begin{align}
\label{E:ConfGraviton}
\delta H_{{\alpha \dt \alpha}} = \bar D_{\dt \alpha} L_\alpha - D_\alpha \bar L{}_{\dt \alpha}
{~~},
\end{align}
defining it as an irreducible superspin-$\tfrac32$ representation. Its component field content is given in table \ref{T:SUGRASpectrum}
\begin{table}[t]
{\renewcommand{\arraystretch}{1.1} 
\begin{center}
\hspace{-5mm}
\begin{tabular}{|c|c|c|c|c|c|c|c|}
\hline
	& conformal vierbein & conformal gravitini & auxiliaries \\
\hline
$H_a$	& $e_m{}^a$	& $\psi_m^\alpha$ & $d_m$ \\
$\Psi^\alpha_i$ & |& $\psi_{mi}^\alpha$ & $y_{m i}$, $t_{abi}^-$, $\rho_{\alpha i}$ \\
\hline
\end{tabular}
\end{center}
} 
\begin{caption}{Component spectrum of superspin-$\tfrac{3}{2}$ and superspin-1 superfields}
\label{T:SUGRASpectrum}
\end{caption}
\end{table}
and consists of the 4D vierbein $e_m{}^a$, the $N=1$ gravitino $\psi_m{}^\alpha$, and the (non-propagating) $U(1)_R$ gauge field $d_m$. Due to the local superconformal symmetry, the trace of the graviton and the gamma-trace of the gravitino are absent and supplied instead by the compensator $G$ via its bottom component $|G|$ and the fermion $\zeta$. In addition, the phase of $G$ is eaten by the $U(1)_R$ gauge field, becoming the massive vector auxiliary field of modified old minimal supergravity.

The remaining superfluous fields, corresponding to two sets of seven fermions $\zeta^i$ and $\zeta_i$ and the two sets of seven scalars in $H_i$ and in the $\bm{7}$ projection of $F_{i j k}$ may naturally be explained as compensators for various symmetries introduced by the $N=1$ formalism but not present in the 11D dynamics. This is the interpretation advocated in \cite{Becker:2016edk}, where it was demonstrated that the (complete non-linear) scalar potential of the component theory is reproduced under the assumption that the lowest components of $H_i$ and $G$ can be gauged to 0 and 1 respectively. Although it was not mentioned in \cite{Becker:2016edk}, the correct normalizations for all spin-0 and spin-1 kinetic terms were also recovered except for those in the $\bm{7}$ of $G_2$.
Based on these facts, it was proposed that the inclusion of the superfields for the additional gravitini resolves the remaining $\bm 7$ problem as well.

In section \ref{S:Linearized}, we will finally prove these claims by explicitly constructing the linearized gravitino couplings and the associated additional gauge symmetries.
We briefly sketch the mechanism here by reviewing the $N=1$ multiplet of a single extra gravitino living purely in four dimensions; details are provided in appendix \ref{S:4DGravitino}.
The 4D, $N=1$ gravitino is described by a spinor superfield $\Psi_{\alpha}$ subject to the linearized gauge transformations \cite{Gates:1979gv}
\begin{align}
\label{E:deltaPsi4D}
\delta \Psi_{\alpha} = \Xi_{\alpha} + D_\alpha \Omega {~},
\end{align}
where $\Xi_{\alpha}$ is chiral and $\Omega$ is an unconstrained complex superfield. The physical content of this multiplet is sketched in table \ref{T:SUGRASpectrum} and consists of the spin-$\tfrac{3}{2}$ gravitino, an auxiliary (non-propagating) complex vector field $y_m \neq \bar y_m$, an auxiliary anti-self-dual antisymmetric tensor $t_{a b}^-$, and an auxiliary fermion $\rho_\alpha$. The large gauge freedom ensures that the gravitino may be taken to be gamma-traceless, while the vector field $y_m$ is a gauge field subject to (complex) abelian gauge transformations. It is important to note that while the dimensions of the bosonic auxiliaries are such that they may appear quadratically in a two-derivative action, the auxiliary fermion must appear as a Lagrange multiplier.\footnote{These features can be understood by considering the field content of 4D $N=2$ conformal supergravity \cite{deWit:1980lyi}. Then $y_m$ corresponds to part of the $SU(2)_R$ gauge field, $t_{ab}^-$ is the bosonic auxiliary field, and $\rho$ corresponds to one of the fermionic auxiliaries $\chi_i$. The additional $N=2$ constituents fill out an $N=1$ vector multiplet.}

In eleven dimensions, we have seven such superfields $\Psi_{\alpha i}$. The fermionic Lagrange multipliers $\rho_{\alpha i}$ can kill the seven extra spinors $\zeta^{\alpha i}$, while $\zeta_{\alpha i}$ can provide the missing gamma-trace of the gravitini. Meanwhile, the complex gauge vectors $y_{m i}$ can eat the remaining 14 extra bosonic degrees of freedom.

As we will show, this is manifested at the superfield level, where we must assign $\Xi$ and $\Omega$ transformations to the prepotentials of section \ref{S:NATH}, and in doing so, it becomes apparent that some of their degrees of freedom may be removed. For example, $\Sigma_{\alpha i}$ must be chosen to transform as a St\"uckelberg field under $\Xi$ as $\delta \Sigma_{\alpha i} = -\Xi_{\alpha i}$, which allows it be gauged away.
These steps lead to the proper 11D spectrum. As we have mentioned, some of the kinetic terms in the $\bm{7}$ require correction terms. These terms (and only these terms) are expected to receive correction, since the complex vectors $y_{mi}$ and the self-dual tensors $t_{ab i}^-$ naturally couple \emph{only} to kinetic terms for the propagating fields; when these auxiliaries are integrated out, the kinetic terms in the $\bm{7}$ are modified.
In appendix \ref{S:4DGravitino}, we demonstrate this explicitly for the linearized action.

\section{Linearized Eleven-dimensional Superspace Action}
\label{S:Linearized}

Because we do not know \emph{a priori} the correct non-linear version of the gravitino transformation \eqref{E:deltaPsi4D} or the corresponding matter field transformations, we cannot immediately couple the gravitino to the full action. One solution to this problem would be to dimensionally reduce 11D superspace to reconstruct the necessary $N\!=\!1$ superfields.
As we are interested only in the linearized action for the moment, a simpler and more expedient approach is to just bootstrap the necessary transformations via the Noether procedure.

We begin by linearizing the prepotentials about a fixed (on-shell) $N=1$ supersymmetric  background. For simplicity, we will take the 4D space-time to be Minkowski (super)space but we will let $Y$ have an arbitrary (but fixed) $G_2$ structure. Backgrounds being on-shell, the $G_2$ 3-form $\varphi_{ijk}(y)$ must be closed and co-closed. We turn off all background flux, so that the background 3-form potential $C_{ijk}$ vanishes. This fixes the background value of $\Phi$ to $\vev{\Phi_{ijk}} = i \varphi_{ijk}$. We take $\vev{V_{i j}}$, $\vev{\Sigma_{\alpha i}}$, and $\vev{\cV^i}$ to vanish to eliminate any space-time flux. Because $G$ is interpreted as a scale compensator, its background sets the Planck scale.
In the normalization of (\ref{E:Kahler}), this corresponds to $\langle G\rangle=1$, which amounts to setting $\langle X \rangle = \theta^2$.

To find the quadratic action describing fluctuations about this background, we replace
\begin{gather}
\Phi_{ijk} \rightarrow i \varphi_{ijk} + \widetilde\Phi_{ijk}{~}, \qquad
V_{ij} \rightarrow \widetilde V_{ij}{~}, \qquad
\Sigma_{\alpha i} \rightarrow \widetilde \Sigma_{\alpha i}{~}, \qquad
X \rightarrow \theta^2 + \widetilde X
\end{gather}
and work to second order in the tilded fields.
The linearized field strengths of the tensor hierarchy are defined as
\begin{subequations}
\begin{align}
\widetilde F &= \frac{1}{2i} (\widetilde \Phi - \bar {\widetilde \Phi}) - \partial \widetilde V{~},\\
\widetilde W_\alpha &= -\frac{1}{4} \bar D^2 D_\alpha \widetilde V + \partial \widetilde \Sigma_\alpha {~},\\
\widetilde H &= \frac{1}{2i} (D^\alpha \widetilde \Sigma_\alpha - \bar D_{\dt \alpha}  \bar{\widetilde\Sigma}{}^{\dt \alpha})
	- \partial \widetilde X {~}, \\
\widetilde G &= -\frac{1}{4} \bar D^2 \widetilde X {~},
\end{align}
\end{subequations}
and these obey the Bianchi identities
\begin{subequations}
\begin{align}
\partial \widetilde F &= \frac{1}{2i} \partial (\widetilde\Phi - \bar{\widetilde\Phi}){~},\\\
-\frac{1}{4} \bar D^2  D_\alpha \widetilde F &= - \partial \widetilde W_\alpha {~},\label{E:LinBI2} \\
\frac{1}{2i} ( D \widetilde W - \bar D \bar {\widetilde W}) &= \partial \widetilde H{~},\\
-\frac{1}{4} \bar D^2 \widetilde H &= -\partial \widetilde G {~}, \\
\bar D_{\dt\alpha} \widetilde G &= 0~.
\end{align}
\end{subequations}
For increased readability, we will drop the tildes from now on. 

For the moment, we neglect the non-abelian gauge prepotential $\cV^i$ \emph{except for its field strength $\cW_\alpha{}^i$}. Then the second-order K\"ahler and Chern-Simons Lagrangians are
\begin{align}
\label{E:LK}
L_K  &= \sqrt{g} \int d^4\theta \,\Big[-\frac{1}{3} G \bar G
	- \frac{1}{9} (F_{\bm 1})^2
	- \frac{1}{12} (F_{\bm{7}})^2
	+ \frac{1}{12} (F_{\bm{27}})^2
	\nonumber \\ & \qquad \qquad	
	+ c\, H_i H_{j} g^{i j}
	- \frac{1}{18} (G + \bar G) \varphi^{i j k} F_{i j k}
	\Big]{~}, \\
L_{CS}
	&= \int d^2\theta \, \left[
	\frac{1}{4} G \,\partial \Phi \wedge \varphi
	-\frac{i}{8} \Phi \wedge \partial \Phi
	+ \frac{1}{8} W^\alpha \wedge W_\alpha \wedge \varphi
	\right.
	\nonumber \\ & \qquad \qquad
	\left.
	+ \frac{i}{4} \partial \Sigma^\alpha \wedge \imath_{\mathcal W_\alpha} \varphi \wedge \varphi
	- \frac{1}{24} \imath_{\mathcal W^\alpha} \varphi \wedge \imath_{\mathcal W_\alpha} \varphi \wedge \varphi
	\right]
	+ \text{c.c.}
\label{E:LCS}
\end{align}
where $c$ is the as-yet-undetermined constant (\ref{E:c}).
To derive this result, one needs the expression \eqref{E:3formMetric} for the Hitchin metric \eqref{E:HitchinMetric} on the space of 3-forms, which leads to the perturbative expansion
\begin{align}
\label{E:RiemMeasureExp}
\frac{\sqrt{g(F)}}{\sqrt{g(F_0)}}
&= 1+ \frac1{18} F_0^{ijk}F_{ijk}
	- \frac1{2} G_0^{ijk , mnp} F_{ijk} F_{mnp}
	+ O(F^3)
\cr
&= 1+ \frac1{18} \varphi^{ijk} F_{\bm 1 ijk}
	-\frac1{2\cdot 18} \left( -\frac43 F_{\bm 1ijk}^2
	-F_{\bm 7 ijk}^2
	+F_{\bm {27}ijk}^2\right)
	+ O(F^3)~.
\end{align}
Here the bold subscripts denote the projection of $F_{ijk}$ onto the corresponding $G_2$ representations with the background $F_{0 ijk} = \varphi_{ijk}$. The explicit form of the projectors is given in \eqref{E:G2Projs}. These Lagrangians are incomplete: We require couplings to the explicit non-abelian gauge prepotential $\cV^i$, the (conformal) supergravity prepotential $H_{\alpha \dt\alpha}$, and the gravitino superfield $\Psi_{\alpha i}$, which we work out sequentially in the next three subsections.

\subsection{The Non-Abelian Gauge Prepotential}

We pause to remind the reader of the distinction between covariant and chiral gauge transformations in $N=1$ superspace. (For a more complete discussion, we refer to the textbooks \cite{Gates:1983nr,Buchbinder:1998qv,Wess:1992cp}.) In a covariant framework, gauge transformations involve unconstrained (usually real) parameters and the transformation rule of a superfield resembles that of its the bottom component. In addition, the superspace derivatives carry a connection, which transforms as a connection should, mirroring the structure of gauge theories in components. Chiral superfields are chiral with respect to the covariant derivative, which is why they may transform with an unconstrained gauge parameter. The gauge prepotential is not explicitly present; rather it is encoded in the covariant derivative and in the chiral superfields themselves.

Until this point, we have used a covariant framework for the non-abelian gauge transformations. The non-abelian gauge prepotential $\cV^i$ was already encoded in the K\"ahler \eqref{E:Kahler} and Chern-Simons actions \eqref{E:NACS} through the covariant derivative \eqref{E:CovDred} and the covariantly chiral superfields. We will need to work with the prepotential explicitly, since it will eventually be required to transform under the hidden supersymmetry. While it could be unpackaged from these objects, the easiest way to restore it in the second-order approximation is to use the Noether method. Since the only $y$-dependent background is $\langle \Phi_{ijk} \rangle= i \varphi_{ijk}$, the only superfields that transform under the linearized non-abelian gauge transformations are $\Phi_{ijk}$ and $\cV^i$ itself. Their transformations are\footnote{One way to motivate these transformations is that in a convenient gauge, the covariantly chiral $\Phi$ is given by $e^{-i \mathscr{L}_{\cV}} \Phi$, which transforms as in \eqref{E:NATHtransformations} with $\tau^i = i (\lambda^i - \bar \lambda)$.}
\begin{align}
\delta \cV^i &= \lambda^i + \bar \lambda^i
{~},{~~~}
\delta \Phi = 2 i \mathscr{L}_{\lambda} \langle \Phi \rangle = -2 \mathscr{L}_{\lambda} \varphi 
{~},{~~~}
\delta \bar\Phi = -2 i \mathscr{L}_{\bar\lambda} \langle \bar \Phi \rangle
	= -2 \mathscr{L}_{\bar\lambda} \varphi
{~},
\label{E:LinNA}
\end{align}
where $\lambda^i$ is chiral and $\bar\lambda^i$ is antichiral.
Here and henceforth, $\Phi$ is chiral with respect to the flat superspace derivatives $D_A$. A straightforward calculation shows that the following terms must be added to $L_K$
\begin{align}
\label{E:LV}
L_K \ni \sqrt{g} \int d^4\theta\, \cV^i \Big[
	i  \,\partial_i (G - \bar G)
	- \frac{1}{3} \varphi^{jkl}
		(\partial_{[i} \Phi_{jkl]} + \partial_{[i} \bar\Phi_{jkl]})
	\Big]{~},
\end{align}
while $L_{CS}$ is gauge-invariant on its own.

\subsection{Coupling to $N=1$ Supergravity}

As we have already mentioned, the superfield $G$ for the space-time 4-form field strength enters the action in a way that suggests it is a conformal compensator. For this reason, we ought to be able to construct a \emph{locally} superconformal quadratic action by coupling to the (conformal) supergravity prepotential $H_{\alpha \dt{\alpha}}$. Local $N=1$ superconformal transformations are encoded in an unconstrained spinor superfield $L_\alpha$ under which $H_{\alpha \dt{\alpha}}$ transforms as \eqref{E:ConfGraviton}. $G$ must also transform as
\begin{align}
\label{E:GXL}
\delta G = -\frac{1}{4} \bar D^2 D^\alpha L_\alpha \quad\Rightarrow\quad
\delta X = D^\alpha L_\alpha + \bar D_{\dt\alpha} \bar L^{\dt\alpha}{~},
\end{align}
which is consistent with the interpretation of $G$ as the conformal compensator of (modified) old minimal supergravity. We review this formulation of 4D, $N=1$ supergravity in appendix \ref{S:OMSG}. Covariantizing the $G \bar G$ term of $L_K$ leads to the linearized (modified) old minimal supergravity Lagrangian (cf.\ eq.\ \eqref{E:OMSG}),
\begin{align}
L_K &\ni \sqrt{g} \int d^4\theta \,\cL_{OMSG} {~}, \nonumber \\
\cL_{OMSG} &=
	-\tfrac13  \bar G G
	+\tfrac {2i}3 (G-\bar G)\partial_a H^a
	- H^a \Box H_a
	\nonumber \\  & \qquad
	+ \tfrac{1}{8} D^2 H_a \bar D^2 H^a
	- (\partial_a H^a)^2
	+ \tfrac1{48}([D_\alpha, \bar D_{\dt \alpha}] H^{\alpha \dt \alpha})^2
{~~}.
\label{E:mOMSGlin}
\end{align}

Because we have set the background value of $G$ to 1, we should no longer refer to the conformal or $U(1)_R$ weights of any quantities. However, since $G$ does not carry engineering dimension, this remains a useful weight. In particular, $L_\alpha$ must carry engineering dimension $d = -\tfrac32$. Then it is not possible to assign linearized $L_\alpha$ transformations to any of the other tensor hierarchy prepotentials on dimensional grounds. (Choosing $\delta \Sigma_{\alpha i} \propto \partial_i L_\alpha$ would violate chirality.)

What about the rest of $L_K$? In order to covariantize the $(G + \bar G) \varphi^{ijk} F_{ijk}$ term in (\ref{E:LK}), we can replace it with the combination
\begin{align}
L_K \ni \sqrt g \int d^4\theta \, \left[
	- \frac{1}{18} \left(G + \bar G
		- \frac{1}{2} [D_\alpha, \bar D_{\dt\alpha}] H^{\dt \alpha \alpha}\right)
		\varphi^{i j k} F_{i j k}
\right]
{~~}.
\end{align}
Under $L_\alpha$ transformations, the part of $F_{ijk}$ involving $\Phi_{ijk}$ drops out, but we are left with the piece involving $\partial_{[i} V_{jk]}$. This can be rearranged to the combination
\begin{align}
\label{E:LKWij}
\delta L_K \ni \sqrt g \int d^4\theta \, \left[
	- \frac{1}{2} \partial_i L^\alpha \,W_{\alpha j k} \varphi^{ijk}
\right]
{~~},
\end{align}
which is invariant under the abelian tensor hierarchy transformations. However, it cannot be countered by introducing a term involving $H_{\alpha\dt\alpha}$.
Similarly, the term \eqref{E:LV} we wish to add to restore gauge invariance under the non-abelian gauge group gives an $L_\alpha$ transformation that cannot be canceled. The most we can do is to covariantize it by replacing \eqref{E:LV} with
\begin{align}
L_K \ni \sqrt{g} \int d^4\theta\, \cV^i \Big[
	\partial_i \left( i  (G - \bar G)
	- 2  \partial_a H^a\right)
	- \frac{1}{3} \varphi^{jkl}
		(\partial_{[i} \Phi_{jkl]} + \partial_{[i} \bar\Phi_{jkl]})
	\Big]{~},
\end{align}
so that the $L_\alpha$ transformation simplifies to
\begin{align}
\label{E:LKWi}
\delta L_K \ni \sqrt g \int d^4\theta\, \left[
	-i \,\partial_i L^\alpha \,\cW_\alpha{}^i
\right]~.
\end{align}
The only other term in $L_K$ that requires covariantization is the $(H_i)^2$ term, since the field strength $H_i$ depends on $\partial_i X$ and $X$ transforms according to (\ref{E:GXL}). However, this again leads to an $L_\alpha$ transformation that cannot be countered by $H_{\alpha\dt\alpha}$ itself.

Finally, eleven-dimensional Lorentz invariance requires that we add to $L_K$ the ``mass'' term for the conformal graviton \cite{Becker:2016edk}
\begin{align}
\label{E:Spin2Mass}
\int d^4\theta \, (\partial_i H^a)^2 =
- \int d^4\theta \, H^a \partial^i \partial_i H_a
\end{align}
normalized to combine with the $H^a \Box H_a$ term in \eqref{E:mOMSGlin} to give the 11D d'Alembertian.\footnote{As we review in appendix \ref{S:OMSG}, the other quadratic terms for $H_a$ in \eqref{E:mOMSGlin} vanish in the superspace Lorentz gauge (\ref{E:LorentzGauge}).} 
The reason it must be \emph{explicitly} included is that while mass terms are often generated by integrating auxiliary fields out of the component action, this term carries spin-2 and none of the auxiliary fields carry spin $>1$. However, this also leads to a $\partial_i L_\alpha$ term.

It does not seem possible to make $L_K$ invariant under $L_\alpha$ transformations \emph{just} by coupling to $H_{\alpha\dt\alpha}$. In fact, we encounter the same problem when taking the $L_\alpha$ transformation of the Chern-Simons Lagrangian $L_{CS}$, which can be written
\begin{align}\label{E:Lalpha.LCS}
\delta L_{CS} = \int d^4\theta \,\left[
\frac{i}{2} \partial L^\alpha \wedge D_\alpha F\wedge \varphi + \text{c.c.}
\right]{~},
\end{align}
again with the same combination of $\partial_i L_\alpha$.

All of these problems have the same solution. While none of the other tensor hierarchy fields may transform under $L_\alpha$, the conformal gravitino superfield can, as it carries the same superspin as $L_\alpha$. In the next section, we will see how this works.

\subsection{The Conformal Gravitino Superfield}
\label{S:Gravitino}

Now we will show how to incorporate the seven missing conformal gravitino superfields $\Psi^\alpha_i$ and construct their couplings to the ``matter'' superfields of the previous section. Being the least familiar multiplet in our construction, we give a self-contained presentation of the 4D, $N=1$ ``matter gravitino'' in appendix \ref{S:4DGravitino}.

The conformal gravitino multiplets have a large linearized gauge transformation
\begin{align}
\label{E:GravitinoXf}
\delta \Psi_{\alpha i} = \Xi_{\alpha i} + {D}_\alpha \Omega_i + 2i \,\partial_i L_\alpha
{~},~~~~i=1,\dots,7
\end{align}
with chiral parameter $\Xi$ and complex unconstrained $\Omega$ describing the irreducible superspin-1 multiplet: At the component level, $\Psi$ contains only spins $\tfrac32$ and $1$. The inclusion of $\partial_i L_\alpha$ is necessary to counter all of the $\partial_i L_\alpha$-dependent terms.\footnote{The factor of $2i$ in \eqref{E:GravitinoXf} just fixes the normalization and phase of $\Psi_{\alpha i}$.} We have already mentioned that the engineering dimension of $L_\alpha$ must be $d=-\tfrac32$. This determines the engineering dimension of $\Psi$ and $\Xi$ to be $d=-\tfrac12$ and that of $\Omega$ to be $d=-1$. We record these in Table \ref{T:Weight.Grav} along with the engineering dimensions of $H_{\alpha \dt\alpha}$ and $\cV$.

\begin{table}[t]
\begin{align*}
{\renewcommand{\arraystretch}{1.3} 
\begin{array}{|c|ccc|cccc|}
\hline
	& H_{\alpha \dt{\alpha}} & \Psi_{\alpha i} & \cV^i & L_\alpha & \Xi_{\alpha i} & \Omega_i & \zeta^i \\
\hline
d  & -1 & -\tfrac12 & -1 & -\tfrac12 & -\tfrac12 & -1 & -1 \\
\hline
\end{array}
}
\end{align*}
\begin{caption}{Engineering dimensions}
\footnotesize
Engineering dimension ($d$) of the graviton, gravitini, and Kaluza-Klein prepotentials and their gauge parameters. Because $\vev{G}=1$, we do not assign conformal and $U(1)_R$ weights.
\label{T:Weight.Grav}
\end{caption}
\end{table}

\subsubsection{Linearized K\"ahler Action}
\label{S:LinKahler}
First, let us see how the Lagrangian $L_K$ may be made fully $L_\alpha$-invariant. The only way to covariantize the mass term is to replace $\partial_i H^a$ with the combination \cite{Linch:2002wg}
\begin{align}
\label{E:CovariantSpin2Mass}
{P}_{\alpha \dt \alpha \,  i} := \tfrac1{2i}\left( \bar {D}_{\dt \alpha} \Psi_{\alpha i} + {D}_\alpha \bar \Psi_{\dt \alpha i} \right)
	- \partial_i H_{{\alpha \dt \alpha}}
{~~}.
\end{align}
This is both $L_\alpha$ and $\Xi$-invariant. Thus the $\Xi L$-covariantized mass term (\ref{E:Spin2Mass}) becomes
\begin{align}
\label{E:Spin2MassCov}
\sqrt{g} \int d^4\theta \, g^{ij} {P}_{a i} {P}_j^a
{~~}.
\end{align}
Similarly, to covariantize the  $(H_i)^2$ term, we can define the combination
\begin{align}
\label{E:CovariantSpin0Mass}
\bm H_i := H_i
	+\tfrac1{2i}\left( {D}^\alpha \Psi_{\alpha i} - \bar {D}_{\dt \alpha} \bar \Psi^{\dt \alpha}_i \right)
{~~}.
\end{align}
This effectively replaces $\Sigma_{\alpha i}$ in the definition of $H_i$ with $\Sigma+\Psi$. This combination can also be made $\Xi$-invariant provided the 2-form gauge superfield shifts
as\footnote{Since these are superfields of the same type (both chiral), the $\Xi$ parameter could be used to eliminate $\Sigma_{\alpha i}$; however, when we go to components, we will use the $\Xi$ transformations to instead impose Wess-Zumino gauge for $\Psi_{\alpha i}$ (cf.\ \S{}\ref{S:WZGravitino}).}
\begin{align}
\delta_\Xi \Sigma_{\alpha i} = - \Xi_{\alpha i}
{~~}.
\end{align}
Countering the residual transformations \eqref{E:LKWij} and \eqref{E:LKWi} requires
\begin{align}
L_K \ni \sqrt g \int d^4\theta \left[
	\frac{1}{2} \Psi^\alpha_i \left(\cW_\alpha^i - \frac{i}{2} W_{\alpha j k} \varphi^{ijk}\right)
	- \frac{i}{4} \varphi^{i j k} \Psi^\alpha_i \partial_j \Psi_{\alpha k}
	+ \text{c.c.}
	\right]
~.	
\end{align}
The first term gives the necessary counter-terms, while the second ensures that the full combination is $\Xi$-invariant.

Including all of the terms, we have found
\begin{align}
\label{E:LKahler.Final}
L_K = \sqrt{g} \int d^4\theta  \,
\left[	\cL_{OMSG}	+ {P}_{a i}^2
	+ c \bm H_i^2
	+ \cL_Y
	+ \cL_{\mathcal V}
\right]
\end{align}
consisting of the following parts:
\begin{itemize}
	\item $\cL_{OMSG}$ is the Lagrangian of (modified) old minimal supergravity in the quadratic approximation \eqref{E:mOMSGlin}, see appendix \ref{S:OMSG}. This covariantizes the $G \bar G$ term in the quadratic action under $L_\alpha$.
	\item The $P^2$ term is the $L$-covariantization (\ref{E:Spin2MassCov}) of the graviton mass term (\ref{E:Spin2Mass}) by the gravitino superfield.
	\item The quadratic term in the 3-form field strength $H_i$ comes from expanding the function $\mathcal H$ in (\ref{E:Kahler}). Here, $H_i$ is covariantized to $\bm H_i$
	\eqref{E:CovariantSpin0Mass} when coupling to the gravitino.
	As discussed in section \ref{S:KahlerReview}, $\mathcal H(x) = 1 + cx +O(x^2)$ for some real constant $c$ (\ref{E:c}).
	Requiring invariance under the $\Omega$ transformations that describe extended supersymmetry will determine $c=-\tfrac14$.
	\item $\cL_Y$ consists of all terms arising from the expansion of the Riemannian volume density on $Y$,
\begin{align}
\label{E:LY}
\cL_Y &=  -\tfrac19 F_{\bm 1ijk}^2
	-\tfrac1{12}F_{\bm 7 ijk}^2
	+\tfrac1{12}F_{\bm {27}ijk}^2
	- \tfrac{1}{18} \big(G + \bar G
		- \tfrac{1}{2} [D_\alpha, \bar D_{\dt\alpha}] H^{\dt \alpha \alpha}\big)
	\varphi^{ijk}F_{\bm 1 ijk}
{~~}.
\end{align}
This contains not only the kinetic terms of the scalars but also the interaction term between the 4D trace of the metric and the volume modulus $\varphi^{ijk}F_{\bm 1 ijk}$ of $Y$.
	\item Finally, $\cL_{\mathcal V}$ is the covariantization under non-abelian gauge transformations (i.e. the internal diffeomorphisms on $Y$)
\begin{align}
\label{E:NonAb}
\cL_\cV = \cV^i \Big[
	\partial_i\left( i   (G - \bar G)
	- 2 \partial_a H^a \right)
	- \tfrac{1}{3} \varphi^{jkl}
		(\partial_{[i} \Phi_{jkl]} + \partial_{[i} \bar\Phi_{jkl]})
	\Big]~.
\end{align}
This gives the component coupling of the Kaluza-Klein gauge field, which appears when one extracts the connection from the covariant derivative \eqref{E:CovDred}.
\end{itemize}

\subsubsection{Quadratic Chern-Simons Action}

Now let us make the Chern-Simons Lagrangian \eqref{E:LCS} invariant as well. It helps to first rewrite it as
\begin{align}
L_{CS}
	&=
	\int d^2\theta \,\left[
	-\frac{i}{8} \Phi \wedge \partial \Phi
	+ \frac{1}{4} G \,\partial \Phi \wedge \varphi
	+ \frac{1}{8} W^\alpha \wedge W_\alpha \wedge \varphi
\right.
\nonumber \\
& \qquad \qquad
\left.
	- \frac{1}{24} \imath_{\cW^\alpha} \varphi \wedge \imath_{\cW_\alpha} \varphi \wedge \varphi
	\right] + \text{c.c.}
	- \frac{1}{2} \int d^4\theta \, \partial H \wedge \imath_\cV \varphi \wedge \varphi~.
\label{E:LCSinitial}
\end{align}
As before, we replace $H_i$ with $\bm H_i$ given by \eqref{E:CovariantSpin0Mass}. This is not actually necessary for $L_\alpha$ invariance (since $H_i$ appears under the $Y$ differential), but it ensures $\Xi$-invariance of this term. Under an $L_\alpha$ transformation, we have already found \eqref{E:Lalpha.LCS}, which can be canceled by adding the additional term
\begin{align}
L_{CS} \ni -\frac{1}{4} \int d^4\theta \,\left[
	\Psi^\alpha \wedge D_\alpha F\wedge \varphi + \text{c.c.}
	\right]~.
\end{align}
Remarkably, the $\Xi$ transformation of this term precisely cancels that of the $(W_\alpha)^2$ term in \eqref{E:LCSinitial}. This feature is quite non-trivial and relies upon the Bianchi identity \eqref{E:LinBI2}. The resulting Lagrangian
\begin{align}
L_{CS}
	&=
	\int d^2\theta \,\left[
	-\frac{i}{8} \Phi \wedge \partial \Phi
	+ \frac{1}{4} G \,\partial \Phi \wedge \varphi
	+ \frac{1}{8} W^\alpha \wedge W_\alpha \wedge \varphi
\right.
	\nonumber \\
	& \qquad \qquad
\left.
	- \frac{1}{24} \imath_{\cW^\alpha} \varphi \wedge \imath_{\cW_\alpha} \varphi \wedge \varphi
	\right]
	+ \text{c.c.}
	\nonumber \\ & \quad
	+ \frac{1}{2} \int d^4\theta \left[ \partial \bm H \wedge \imath_\cV \varphi \wedge \varphi
	- \frac{1}{2} \Psi^\alpha \wedge D_\alpha F\wedge \varphi
	- \frac{1}{2} \bar\Psi_{\dt \alpha} \wedge \bar D^{\dt\alpha} F\wedge \varphi \right]
\label{E:LCSfin.v1}
\end{align}
is invariant under both $L_\alpha$ and $\Xi_\alpha$. This can be rewritten with
explicit indices as
\begin{align}
L_{CS}
	&= \int d^2\theta \, \left[
	-\frac{i}{288} \epsilon^{ijklmnp} \,\Phi_{ijk} \partial_l \Phi_{mnp}
	+ \frac {1}{24} G \,\tilde \varphi^{ijkl} \partial_i \Phi_{jkl}
\right.
	\nonumber \\ & \qquad \qquad
\left.	
	+ \frac{1}{32}\tilde \varphi^{ijkl} W^{\alpha}_{ij}  W_{\alpha kl}
	+\frac{1}{4} \mathcal W^{\alpha i}  \mathcal W_{\alpha}^j \, g_{ij}
\right]
	+ \text{c.c.}
	\nonumber \\ & \quad
	+ \int d^4\theta \, \left[
	- \partial_i \bm H_j \, \varphi^{ij}{}_k \cV^k
	- \frac{1}{24} \Psi^\alpha_i D_\alpha F_{jkl} \,\tilde\phi^{ijkl}
	- \frac{1}{24} \bar\Psi_{\dt\alpha i} \bar D^{\dt\alpha} F_{jkl} \,\tilde\phi^{ijkl}
	\right]~.
\label{E:LCSfin.v2}
\end{align}
The F-term contains both the $G_2$ superpotential $\Phi\partial \Phi$ and the kinetic terms for the vector multiplets.

\subsubsection{Invariance under Extended Supersymmetry}
\label{S:ExtendedSUSY}

We have not yet discussed the $\Omega$ part of the linearized gravitino transformation. Requiring gauge invariance of the combined K\"ahler-Chern-Simons action must fix the linearized $\Omega$ transformations of the other fields. These turn out to be
\begin{subequations}
\begin{align}
\delta \Phi_{ijk} &= -\frac{i}{2} \widetilde\varphi_{ijkl} \bar D^2 \bar\Omega^l{~}, \\
\delta V_{ij} &= -\frac{i}{2} \varphi_{ijk} (\Omega^{k} - \bar \Omega^{k}){~}, \\
\delta \cV^i &= -\frac{1}{2} (\Omega^i + \bar \Omega^i){~},
\end{align}
\end{subequations}
with $\Sigma_{\alpha i}$, $X$, and $H_{\alpha \dt\alpha}$ invariant. While they can be determined directly by requiring invariance of the action, the structure of the transformations (up to normalization) can be determined purely on the grounds of symmetry and a few observations. These transformations also fix the constant in (\ref{E:c}) to $c = \cH'(0) = -\tfrac14$.

Let's briefly motivate why the structure of the $\Omega$ transformations must be of this form. Since they are linearized, they may contain only the background metric and $\varphi_{ijk}$ in addition to $\Omega$ and its derivatives. The engineering dimension forbids any derivatives from appearing in $\delta V_{ij}$ and $\delta \cV^i$ while $\delta \Phi$'s dimension and chirality permit only $\bar D^2$. To determine that $\bar D^2$ acts on $\bar\Omega$ rather than $\Omega$ (or both), one must recall that $\Omega$ appears in its defining transformation \eqref{E:GravitinoXf} under a $D_\alpha$, so it can be shifted by an anti-chiral superfield. This is a superfield version of a gauge-for-gauge symmetry, and it is necessary so that the physical content of $\Omega$ (and $\Xi$) are precisely enough to adopt a proper Wess-Zumino gauge condition for $\Psi$. (We will discuss the physics of this in the next section.) This gauge-for-gauge symmetry is manifestly maintained in $\delta\Phi$ only for $\bar D^2 \bar\Omega$, while for $\delta V_{ij}$ and $\delta \cV^i$ the shift in $\bar\Omega$ can be countered by a certain non-abelian $\lambda^i$ transformation combined with an abelian $\Lambda_{ij}$ transformation. Finally, the phase in $\delta \Phi_{ijk}$ and the requirement that the imaginary part of $\Omega$ be used for $\delta V_{ij}$ and the real part for $\delta \cV^i$ can be determined by requiring invariance under space-time parity.

\subsection{Assimilation and Summary}
\label{S:Assimilation}

We now collect all the terms we have worked out for the linearized eleven-dimensional supergravity action. 
This section summarizes our main result.

The complete action for eleven-dimensional supergravity (to quadratic order in fields) is given by the sum of the K\"ahler \eqref{E:LKahler.Final} and Chern-Simons actions \eqref{E:LCSfin.v2}. As we have emphasized, this action is invariant under a large set of superspace gauge
transformations:\footnote{All
fields transform under the full supergravity gauge group, but we are presenting only the non-vanishing linearized transformations here. The non-linear corrections are important for the full action, but we defer this to future work.
} 
\begin{enumerate}
\item The abelian tensor hierarchy transformations \cite{Becker:2016xgv} (cf.\ \ref{E:NATHtransformations}) \begin{subequations}
	\label{E:ATHlin}
	\begin{align}
	\delta_{ATH} \Phi_{ijk} &=
		3\partial_{[i} \Lambda_{jk]}
	\\
	\delta_{ATH} V_{ij} &=
	\tfrac1{2i}\left(\Lambda_{ij} - \bar \Lambda_{ij} \right)
		- 2\partial_{[i} U_{j]}
	\\
	\delta_{ATH} \Sigma_{\alpha i}&=
		-\tfrac14 \bar {D}^2 {D}_\alpha U_i
		+ \partial_i \Upsilon_\alpha
	\\
	\delta_{ATH} X &=
		\tfrac1{2i}\left({D}^\alpha \Upsilon_\alpha - \bar {D}_{\dt \alpha} \bar \Upsilon^{\dt 	\alpha} \right)
	{~~}.
	\end{align}
	\end{subequations}
\item The non-abelian gauge transformations (i.e. the internal diffeomorphisms) with chiral parameter $\lambda^i$ \cite{Becker:2016xgv} (cf. \eqref{E:LinNA})
	\begin{subequations}
	\label{E:NAlin}
	\begin{align}
	\delta_\lambda \mathcal V^i &= \lambda^i + \bar \lambda^i{~}, \\
	\delta_\lambda \Phi_{ijk} &= -6 \,\partial_{[i} ( \varphi_{j k] l} \lambda^l ){~~}.
	\end{align}
	\end{subequations}
	The transformation of $\Phi_{ijk}$ can be interpreted as a certain abelian $\Lambda_{ij}$ transformation. This means one can choose to define a covariantized non-abelian
	transformation that is often easier to work with:
	\begin{subequations}
	\begin{align}
	\delta'_\lambda \mathcal V^i &= \lambda^i + \bar \lambda^i{~}, \\
	\delta'_\lambda V_{i j} &= -i \varphi_{ijk} (\lambda^k - \bar \lambda^k)~.
	\end{align}
	\end{subequations}
\item The superconformal gravitino transformations with chiral parameter $\Xi_{\alpha i}$ and complex parameter $\Omega^i$
\begin{subequations}
\label{E:Grinolin}
\begin{align}
\delta_{\Xi\Omega} \Psi_{\alpha i} &= \Xi_{\alpha i} + D_\alpha \Omega_i
\\
\delta_{\Xi\Omega} \Phi_{ijk} &= \tfrac1{2i} \tilde \varphi_{ijkl} \bar D^2 \bar \Omega^l
\label{E:Grinolinb}
\\
\delta_{\Xi\Omega} V_{ij} &= \tfrac1{2i} \varphi_{ijk} (\Omega^k -\bar \Omega^k)
\\
\delta_{\Xi\Omega} \Sigma_{\alpha i}&= -\Xi_{\alpha i}
\\
\delta_{\Xi\Omega} \mathcal V^i &= -\tfrac12(\Omega^i + \bar \Omega^i)
{~}.
\end{align}
\end{subequations}
\item The local superconformal transformations with parameter $L_\alpha$ under which
\begin{subequations}
\label{E:ConfSGlin}
\begin{align}
\delta_L H_{\alpha \dt \alpha} &= \bar D_{\dt \alpha} L_\alpha - D_\alpha \bar L_{\dt \alpha}
\\
\delta_L X &= D^\alpha L_\alpha + \bar D_{\dt \alpha}  \bar L^{\dt \alpha}
\\
\delta_L \Psi_{\alpha i} &= 2i \,\partial_i L_\alpha
{~~}.
\end{align}
\end{subequations}
\end{enumerate}

We present the complete linearized action invariant under these transformations in terms of D- and F-term integrals
\begin{align}
\label{E:FinalAction}
S= \frac1{\kappa^2} \int d^4x \,d^7y \int d^4 \theta \,L_D
	+\frac1{\kappa^2} \int d^4x \,d^7y \, \left[ \int d^2 \theta \,L_F +\textrm{h.c.}\right]
\end{align}
in 4D, $N=1$ superspace extended to $Y$.
Putting together the pieces of the previous sections, we find the quadratic Lagrangians to be
\begin{subequations}
\label{E:Final}
\begin{align}
L_D &=
	- H^a \Box H_a
	+ \frac{1}{8} D^2 H_a \bar D^2 H^a
	- (\partial_a H^a)^2
	+ \frac1{48}([D_\alpha, \bar D_{\dt \alpha}] H^{\alpha \dt \alpha})^2
	\cr	& 
	- \frac{1}{3} \bar G G
	+\frac{2i}{3} (G-\bar G)\partial_a H^a
\label{E:FinalOMSG}
	\\ &	
	- \!\frac1{18} \big( G + \bar G
	- \frac{1}{2} [D_\alpha , \bar D_{\dt \alpha}] H^{\alpha \dt \alpha}\big) \varphi^{ijk} F_{ijk}
	- \frac1{9} F_{\bm 1ijk}^2
	- \!\frac1{12}F_{\bm 7 ijk}^2
	+ \frac1{12}F_{\bm {27}ijk}^2
\label{E:FinalRadionFF}
	\\ & 
	-\frac12 \Big[\partial_i H_{{\alpha \dt \alpha}}
	- \frac1{2i} (\bar D_{\dt \alpha} \Psi_{\alpha i}
	+ D_\alpha \bar \Psi_{\dt \alpha i} )\Big]^2
	- \frac{1}{4} \Big[ H_i + \frac1{2i} (D^\alpha \Psi_{\alpha i} - \bar D_{\dt \alpha} \bar \Psi^{\dt \alpha}_i )\Big]^2
\label{E:Gravitino2}
\\ & 
	+\frac{1}{2} \Big\{ \Psi^\alpha_i \Big[
	\mathcal W_\alpha^i
	-\frac{i}{2} \varphi^{ijk} (\partial_j \Psi_{\alpha k} + W_{\alpha jk} )
	- \frac{1}{12} \tilde \varphi^{ijkl} D_\alpha F_{jkl}
	\Big]
	+\textrm{h.c.} \Big\}
\label{E:Current}
	\\ & 
	+ \cV^i \Big[
	i  \,\partial_i (G - \bar G)
	- 2 \,\partial_i \,\partial_a H^a
	- \frac{1}{3} \varphi^{jkl}
		(\partial_{[i} \Phi_{jkl]} + \partial_{[i} \bar\Phi_{jkl]})
	\cr & \hspace{56mm}
	- \varphi_{i}{}^{j k} \partial_j \Big(
	H_k + \frac1{2i} (D^\alpha \Psi_{\alpha k} - \bar D_{\dt \alpha} \bar \Psi^{\dt \alpha}_k)
	\Big)
	\Big]
\label{E:Gauging}
\\
\label{E:FtermFinal}
L_F &= -\frac{i}{288} \epsilon^{ijklmnp} \,\Phi_{ijk} \partial_l \Phi_{mnp}
	+ \frac {1}{24} \tilde \varphi^{ijkl} \,G \,\partial_i \Phi_{jkl}
	\cr & \hspace{65mm}
	+ \frac{1}{32}\tilde \varphi^{ijkl} W^{\alpha}_{ij}  W_{\alpha kl}
	+\frac{1}{4} \, g_{ij} \,\mathcal W^{\alpha i}  \mathcal W_{\alpha}^j~.
\end{align}
\end{subequations}
We have organized the terms as follows:
\begin{description}
	\item[(\ref{E:FinalOMSG})] This is the action of linearized old minimal supergravity (\ref{E:mOMSGlin}).
	\item[(\ref{E:FinalRadionFF})] These are the terms (\ref{E:LY}). They describe the ``radion coupling'' between the 4D and 7D volume terms and $F_{ijk}$ kinetic terms.
	\item[(\ref{E:Gravitino2})] Quadratic gravitino terms and ``mass'' terms for the prepotentials $H^a$ and $X$ of modified old minimal supergravity have the sum-of-squares form as follows from $L$-invariance.
	\item[(\ref{E:Current})] This includes the gravitino current, describing the linear couplings of $\Psi$ to the Kaluza-Klein gauge fields, the $Y$ components of the 3-form, and the ``mass'' of the gravitino.
	\item[(\ref{E:Gauging})] Kaluza-Klein gauge field couplings are needed to covariantize the linearized diffeomorphisms of $Y$. In a more covariant description, these couplings are hidden in the covariant derivative $\mathcal D$ (\ref{E:CovDred}). 
	\item[(\ref{E:FtermFinal})] The F-term contains the $G_2$ superpotential $\Phi \partial \Phi$ \cite{Becker:2014rea} and gauge kinetic terms. The gauge symmetry (\ref{E:Grinolinb}) (or more precisely \eqref{E:NoGrav7}) explains the consistency of this superpotential for the first time.
\end{description}

This action and its gauge transformations are the main result of this paper.
We have endeavored to present it in a way that motivates the roles of the myriad parts and how they relate to one another under.
Pragmatically, the presentation of the foregoing sections can be skipped, and the claim that the action (\ref{E:FinalAction}|\ref{E:Final}) is invariant under the transformations (\ref{E:ATHlin}|\ref{E:ConfSGlin}) can be checked directly.

In the next section, we project our action to components to demonstrate explicitly that this is indeed the superspace representation of the linearized action of eleven-dimensional supergravity.

\section{Components}
\label{S:Components}

We now want to confirm that the Lagrangian (\ref{E:Final}) produces the correct component action of eleven-dimensional supergravity on $M= \mathbf R^4 \times Y$ and elucidate the required auxiliary field mechanisms.
The part of the Lagrangian with all derivatives and polarizations along $\mathbf R^4$ corresponds to 4D, $N=1$ supergravity, while the part with all derivatives and polarizations along $Y$ (corresponding to the scalar potential in 4D) was demonstrated at the fully non-linear level in \cite{Becker:2016edk}. Thus, we will mainly be interested in the mixed part at the linearized level.

To compare to the quadratic approximation of the eleven-dimensional component action of reference \cite{Becker:2014uya}, we rewrite that result in terms of $G_2$ representations.
Using various $G_2$ identities which we collect in appendix \ref{S:G2} leads to the Lagrangian
\begin{subequations}
{\renewcommand{\arraystretch}{1.7} 
\label{E:TheAnswer}
\begin{align}
\kappa^2 L^{(2)} = {}&{}
	-\frac18 (\partial_c h_{ab})^2
	+ \frac14 (\partial^b h_{ab})^2
	+ \frac18 (\partial_a h)^2
	+ \frac14 h \partial^a \partial^b h_{ab}
\label{E:Spin2}
\\
&	-\frac18 (\partial_i h_{ab})^2 +\frac18 (\partial_i h)^2
\label{E:Mass}
\\
&	+\frac1{36} \left( \partial^a \partial^b h_{ab} - \Box h \right)\left(\varphi^{ijk}F_{\mathbf 1 ijk}\right)
\label{E:Radion}
\\
\label{E:MixedPotential}
&		-3 h \partial^i (\tau_1)_i
\\
\label{E:7s}
&	-\frac1{8} (\mathcal F_{ab}^i)^2
		+\frac12 \partial_a h^{ab} \partial_i \mathcal A^i_b
		-\frac12 \partial^a h \partial_i \mathcal A^i_a
\\
\label{E:Multi-Maxwell}
&	-\frac1{96} F_{abcd}^2
	-\frac1{24} F_{abc\, i}^2
	-\frac1{16} \left( F_{\mathbf 7\,ab\, ij}^2
	+F_{\mathbf{14}\, ab\, ij}^2 \right)
\\
\label{E:ScalarKinetic}
&		+\frac1{18} (\partial_a \phi_{\mathbf 1 ijk})^2 - \frac1{24} (\partial_a \phi_{\mathbf {27}ijk})^2
		-\frac1{24} \left( F_{\mathbf 1\, a\, ijk}^2
		+F_{\mathbf 7\, a\, ijk}^2
		+F_{\mathbf {27}\, a\, ijk}^2\right)
\\
\label{E:TorsionClasses}
&	+\frac{21}{16} \tau_0^2
			+ 15\tau_1^2
			-\frac18 \tau_2^2
			-\frac1{24} \tau_3^2
\\
\label{E:CClasses}
&		-\frac{7}{4} \sigma_0^2
		- 9 \sigma_1^2
		-\frac1{24} \sigma_3^2
\end{align}
}%
\end{subequations}
The notation and structure of this action are as follows:
\begin{description}
	\item[(\ref{E:Spin2})] is the linearized four-dimensional Einstein-Hilbert action (\ref{E:OMSGComponents}).
	\item [(\ref{E:Mass})] extends the derivatives on the linearized metric to $Y$. From the point of view of four-dimensional compactifications, these look like mass terms for the graviton.
	\item[(\ref{E:Radion})] gives the ``radion coupling'', that is, the coupling between the graviton and the volume modulus of $Y$.
	\item[(\ref{E:MixedPotential})] gives the analogous coupling with $Y$ derivatives. The torsion class $\tau_1$ is the $\mathbf 7$-projection (\ref{E:TorsionClassesDef}) of $\partial_{[i} F_{jkl]}$, the differential of the fluctuation around the $G_2$-structure 3-form.
	\item[(\ref{E:7s})] are the kinetic terms and spin-2 mixing terms of the Kaluza-Klein gauge field (mixed components of the frame) $\mathcal A_a^i = - e_a{}^i$. At quadratic order, this is the entire contribution to the action from this field. (There is no $\varphi_{ijk}\mathcal F_{ab}^{i} F_{\mathbf 7}^{ab\, jk}$ term; the combination that appears instead is the cubic term $C_{ijk}\mathcal F_{ab}^{i} F_{\mathbf 7}^{ab\, jk}$, which we ignore in the quadratic approximation.)
	\item[(\ref{E:Multi-Maxwell})] are the covariantized kinetic terms for the 3-form $C_{abc}$, seven 2-form $C_{ab i}$, and twenty-one 1-form $C_{a ij}$ components of the eleven-dimensional 3-form $C_{\bm{abc}}$.
	\item[(\ref{E:ScalarKinetic})] gives the kinetic terms for the scalars. The first two terms are those for the 28 metric scalars $g_{ij}$ written in terms of the differential of the fluctuations around the $G_2$-structure 3-form. (These have no $\mathbf 7$ part.) The remaining terms are the projections of the covariantized kinetic term of the 35 scalars $C_{ijk}$.
	\item[(\ref{E:TorsionClasses})] is the Einstein-Hilbert term $\tfrac12\sqrt{g}R(g)$ on $Y$ written in terms of torsion classes (\ref{E:TorsionClassesDef}) using a result due to Bryant \cite{Bryant2003}.
From the four-dimensional point of view, these resemble potential terms.
	\item[(\ref{E:CClasses})] gives the Maxwellian contribution $-\tfrac1{4 \cdot 4!} \sqrt{g}F_{ijkl}^2$ on $Y$ in a form analogous to (\ref{E:TorsionClasses}). These also look like potential terms from the four-dimensional perspective.
\end{description}

We now confirm that our Lagrangian (\ref{E:Final}) reproduces the bosonic components (\ref{E:TheAnswer}), beginning with the parts that have already been verified.
The first line (\ref{E:Spin2}) is the linearized gravity action. 
It comes from the component projection (\ref{E:OMSGComponents}) of linearized old minimal supergravity which we review in appendix \ref{S:OMSG}.
The last two lines (\ref{E:TorsionClasses}) and (\ref{E:CClasses}) have already been confirmed at the non-linear level in reference \cite{Becker:2016edk}.
It was also mentioned there that the first term in (\ref{E:Multi-Maxwell}) was used to fix the $G$-dependence of the volume functional (\ref{E:Kahler}).
The coefficient of the mass term for the spin-$2$ field (the first one in line \eqref{E:Mass}) must be the same as the first term in the pure gravity sector (\ref{E:Spin2}).
This becomes clear when working in transverse-traceless gauge wherein the statement amounts to one of 11D Lorentz invariance.
(Lorentz invariance guarantees that the trace mass works as well, but we will discuss it in more detail after we have understood the terms in the $\bm 7$ representation.)

The kinetic terms for fields in the $\bm {14}$ and $\bm{27}$ are particularly simple to work out as these are not corrected by integrating out any auxiliary fields. (The only vector auxiliaries are in the $\bm{1}$ and $\bm{7}$ representations of $G_2$.)
As an example, we check the $\bm {27}$.
Consider the components that would result from the terms (\ref{E:FinalRadionFF} $=$ \ref{E:LY}) in the superfield Lagrangian. (These arose from the expansion (\ref{E:RiemMeasureExp}) of the Hitchin functional and the superspace volume measure.)
To obtain the coefficient of $F_{\bm {27}a \,ijk}^2$, we need only take into account a factor of $-\tfrac12$ that results from the component projection\footnote{The fermionic integral of the square of one of the field strengths $F$ gives
$\int d^4 \theta F^2 = -\tfrac12(\partial_a F)^2 -\tfrac12 F_a^2  + \dots$
where the ellipses stand for fermionic terms and auxiliary fields.
On the right-hand side $F$ stands for the $\theta\to 0$ component of the superfield $F$ and $F_a$ is the $\theta\to 0$ projection of $-\tfrac14 (\sigma_a)^{\dt \alpha \alpha} [D_\alpha , \bar D_{\dt \alpha} ]F$. 
}
to get $-\tfrac12\times \tfrac1{12} = -\tfrac1{24}$. 
Since the $(\partial_a \phi_{\bm {27}ijk})^2$ term is the partner of this, it gets the same factor.

Next, consider the terms $\phi_{\bm 1}$ and $F_{\bm 1 a}$ in (\ref{E:Multi-Maxwell}) and (\ref{E:ScalarKinetic}).
The first of these comes out correctly from the expansion (\ref{E:LY}) with the same factor: $-\tfrac12 \times (-\tfrac 19) = \tfrac1{18}$.
Since $F_{\bm 1 aijk}$ is the pseudoscalar partner of $\partial_a \phi_{\bm 1}$, it too will come with a factor of $\tfrac 1{18}$.
However, this field strength couples to the conformal supergravity auxiliary field $d^a$ (\ref{E:OMSGComponents}) giving a correction
\begin{align}
\label{E:SingletKineticCorrection}
~\hspace{-5mm}
\tfrac43 d_a^2 + \tfrac1{18} d^a \varphi^{ijk} F_{a\, ijk} =
	\tfrac43 \left( d_a + \tfrac1{48} \varphi^{ijk} F_{a\, ijk}\right)^2
	- \tfrac43 \cdot \tfrac{42}{48^2} F_{\mathbf 1\, a\, ijk}^2
~\to~
- \tfrac7{72} F_{\mathbf 1\, a\, ijk}^2 ,
\end{align}
where we used $(\varphi^{ijk}F_{a\, ijk})^2= (\varphi^{ijk}F_{\mathbf 1 a\, ijk})^2 = 42 F_{\mathbf 1 a\, ijk}^2$ (\ref{E:contractions}).
This changes $\tfrac 1{18} \to \tfrac 1{18} - \tfrac7{72} = -\tfrac1{24}$, which is the correct coefficient.
We thus reproduce the terms in (\ref{E:Multi-Maxwell}) and (\ref{E:ScalarKinetic}).

Next, we will look at the radion couplings (\ref{E:Radion}) which come from the first term in (\ref{E:FinalRadionFF}). 
In the Wess-Zumino gauge adopted in \S{}\ref{S:SGWZ},
\begin{align}\label{E:3.37aComp}
\tfrac{1}{36}\int d^4\theta \,[D_\alpha , \bar D_{\dt \alpha}] H^{\alpha \dt \alpha} \varphi^{ijk} F_{ijk} &= \tfrac1{36} \partial^a \partial^b h^{\text{spin-2}}_{ab} \varphi^{ijk} F_{ijk} + \cdots
\end{align}
gives one of these couplings directly for the traceless part of the metric (\ref{E:ConfMetric}),
and
\begin{align}\label{E:3.37bComp}
- \tfrac1{18}  \int d^4 \theta \, \left( G + \bar G\right) \varphi^{ijk} F_{ijk}
&=
- \tfrac1{48}  \Box h \varphi^{ijk} F_{ijk}
- \tfrac1{8} \varphi^{ijk} d_{ij}  \partial_k h
+ \cdots
\end{align}
gives couplings involving the trace of the metric, $h = \frac{8}{3} \textrm{Re\,} G$ \eqref{E:Metric}.
Together, these give the correct radion coupling (\ref{E:Radion}) in the basis in which the spin-0 part of the metric is separated out:
$\tfrac1{36} (\partial^a \partial^b h_{ab} - \Box h) =\tfrac1{36} \partial^a \partial^b h^{\text{spin-2}}_{ab} - \tfrac1{48} \Box h$.
The elided term in \eqref{E:3.37aComp} involves $d_a$, which we have already accounted for, while those in \eqref{E:3.37bComp} involve the auxiliary fields of $G$, $V_{ij}$ and $\Phi_{ijk}$ already included in the analysis of \cite{Becker:2016edk}, where they were shown to generate {\it e.g.} the correct normalization of the 3-form kinetic term. 
This leaves the term that involves the $\bm 7$-projection $d^k := \tfrac16\varphi^{ijk} d_{ij}$ (\ref{E:7components2}) of the auxiliary field of $V_{ij}$. 
Additional terms involving this projection come from the $F^2$ terms in (\ref{E:FinalRadionFF}) and the $\tilde \varphi WW$ the term in (\ref{E:FtermFinal})
which were important in \cite{Becker:2016edk} in obtaining the correct $\tau_1$ contribution to the scalar potential.
Here they will contribute to the trace mass in (\ref{E:Mass}) and the mixed term (\ref{E:MixedPotential}).
Explicitly, we find
\begin{align}
&- \tfrac1{8} \varphi^{ijk} d_{ij}  \partial_k h 
-2 \varphi^{ijk} d_{ij} (\tau_1)_k
+\tfrac14 \tilde \varphi^{ijkl} d_{ij} d_{kl}
\cr &=
- \tfrac 3{4}d^k  \partial_k h 
-12 d^k (\tau_1)_k
-6d_k^2
+ \bm {14}\textrm{-term}
\end{align}
where we used (\ref{E:2formSquare}). Integrating out $d_k$ gives
\begin{align}
\label{E:traceJunk1}
\tfrac 32 \cdot \left(\tfrac 1{8}\right)^2 (\partial_i h)^2 - \tfrac 34 h \partial^i (\tau_1)_i~.
\end{align}

There are two other sources contributing to such terms.
The first is due to the $\bm 7$ projections $f^i$ (\ref{E:7components3}) of the $\Phi$ auxiliaries $f_{ijk}$. 
The square of these terms comes from the $F^2$ part of the K\"ahler action with the linear terms coming from the Chern-Simons terms making up the first line of (\ref{E:FtermFinal})
\begin{align}
-\tfrac1{24} |f_{ijk}|^2
+\tfrac1{24} \cdot \tfrac 38 \tilde \varphi^{ijkl} \mathrm{Re}f_{ijk} \partial_l h 
= 
-\tfrac12 (\mathrm{Re} f_k)^2 + \tfrac 3{16} \mathrm{Re} f^k \partial_k h  
+\dots
\end{align}
where the ellipses stand for terms irrelevant to this calculation. 
Integrating out $\mathrm{Re} f^i$ gives 
\begin{align}
\label{E:traceJunk2}
\left(\tfrac 3{8}\right)^2 (\partial_i h)^2 - \tfrac 94 h \partial^i (\tau_1)_i 
{~~}. 
\end{align}

The second contribution comes directly from the 2-form field strength $H_i$ in (\ref{E:Gravitino2}). 
This term gives only a trace mass correction
$-\tfrac 14 \int d^4\theta (\partial_i X)^2 
= -\tfrac12 \cdot \left(\tfrac38\right)^2 (\partial_i h)^2$.
Adding this to (\ref{E:traceJunk1}) and (\ref{E:traceJunk2}), we find 
\begin{align}
\left(\tfrac 3{8}\right)^2 \left[ \tfrac 16 +1-\tfrac12   \right] (\partial_i h)^2 
+\left[ - \tfrac 34- \tfrac 94 \right] h \partial^i (\tau_1)_i
=
\tfrac3{32} (\partial_i h)^2 - 3 h \partial^i (\tau_1)_i
\end{align}
giving the correct mixed term (\ref{E:MixedPotential}).
Recombining with the traceless part of the metric 
$-\tfrac18(\partial_i h_{ab})^2 + \tfrac18 (\partial_i h)^2 = 
-\tfrac18(\partial_i h^{\text{spin-2}}_{ab})^2 + (-\tfrac 1{32}+ \tfrac 4{32}) (\partial_i h)^2$
shows that this is also the correct trace mass (\ref{E:Mass}).

\subsection{Component Fields in the 7 Representation of $G_2$}
\label{S:7}

Now we finally come to the analysis of the recalcitrant terms in the $\bm 7$ representation of $G_2$.
For this, we apply the general analysis of the gravitino and its compensators worked out in detail in appendix \ref{S:4DGravitino}.
To understand what this general analysis implies for eleven-dimensional supergravity, we compare the Lagrangians (\ref{E:GravitinoLag}) and (\ref{E:Final}). This gives the coefficients
\begin{align}
\label{E:abc11D}
a = -\frac14
{~~},~~~
b= \frac12
{~~},~\textrm{and}~~~
c= - \frac14
\end{align}
for linearized eleven-dimensional supergravity.
Substituting into the component result (\ref{E:CompRes}), gives (using \ref{E:squares})
\begin{align}
\label{E:7Kinetic}
L_{\left(-\tfrac14,\tfrac12,\tfrac12\right)} &=
	\tfrac1{16} \tilde H^i_a \tilde H_i^a
	-\tfrac14 F^a_i F^i_a
	-\tfrac 38 F_i^{ab} F^i_{ab}
	-\tfrac 1{8} \mathcal F_i^{ab} \mathcal F^i_{ab}
\cr	&=
	-\tfrac1{24} (F_{abc i})^2
	-\tfrac1{24} (F_{\bm 7 a ijk})^2
	-\tfrac 1{16} (F_{\bm 7ab ij})^2
	-\tfrac 1{8} (\mathcal F^i_{ab})^2
\end{align}
These are the correct coefficients of the $\bm 7$-projection of the component theory as
found in lines (\ref{E:Multi-Maxwell}), (\ref{E:ScalarKinetic}), and (\ref{E:7s}).
Note that the cancellation of $(\partial_a F_i)^2\propto (\partial_a F_{\bm 7 ijk})^2$ is important to recover the eleven-dimensional theory, since the $\bm 7$ projection of $F_{ijk}$
does not correspond to any physical field.

At this point we have verified all the components in (\ref{E:TheAnswer}) except for the $\partial h\partial \mathcal A$ terms in (\ref{E:7s}). These terms are not subtle, coming directly from (\ref{E:Gauging}): The third $\mathcal V\partial \Phi$ term contributes only to the potential \cite{Becker:2016edk} and the first two integrate to 
\begin{align}
\int d^4\theta \, \cV^i \partial_i \left[
	i  (G - \bar G)
	- 2  \partial_a H^a\right]
&= \partial_i \mathcal A^i_a \left[ \tfrac12 \partial_b h^{\text{spin-2} \,ab}
	- \partial^a \mathrm{Re}G 
\right]
\cr&
= 
\tfrac12 \partial_i \mathcal A^i_a\left[ \partial_b h^{ab} -\partial^a h\right]
{~}.
\end{align}
This completes the verification that the superspace action (\ref{E:FinalAction}) reproduces the bosonic action (\ref{E:TheAnswer}) of eleven-dimensional supergravity in the quadratic approximation.

\section{Conclusions and Outlook}
\label{S:CandO}

We have extended the construction of the embedding of eleven-dimensional supergravity into 4D, $N=1$ superspace of ref. \cite{Becker:2016edk} to quadratic order in the gravitino superfield $\Psi_{\alpha i}$.
This extension is needed to prove that the spectrum is represented faithfully in terms of 4D, $N=1$ superfields, and to show that the dynamics of the eleven-dimensional components is that required by eleven-dimensional Poincar\'e invariance.
The result is that the action is given by (\ref{E:FinalAction}) in terms of the Lagrangian (\ref{E:Final}).
In addition to being manifestly invariant under local 4D, $N=1$ supergravity transformations (\ref{E:ConfSGlin}), the complete action is invariant under the tensor hierarchy gauge transformations (\ref{E:ATHlin}) and (\ref{E:NAlin})
and the extended supersymmetry transformations (\ref{E:Grinolin}).
This level of approximation ({\it i.e.}\ to quadratic order in the gravitini) suffices to demonstrate the consistency of this superspace description of eleven-dimensional supergravity. (In particular, it realizes the linearized gauge symmetry and associated compensator mechanism advocated in \cite{Becker:2016edk}.)

Additionally, we expect it to be adequate for most applications.
For example, already at this level, the conformal graviton propagator and all other superspace Feynman rules needed for perturbative calculations can be deduced as was done for five-dimensional supergravity in ref. \cite{Buchbinder:2003qu}.

There are three directions in which we are currently extending this analysis.
The first is that we would like to complete the quadratic gravitino action to all orders in the remaining fields.
(The analogous $O(H_a, \Psi)$ but all orders in remaining fields has already been worked out and will be presented in a separate article \cite{nonlinear}.)
Secondly, we would like to construct the terms cubic and higher in the gravitino multiplet.
Although this may initially appear a daunting task, current results suggest that it is possible to construct fields strengths invariant under the $\Xi$ transformations. As the K\"ahler part of the action is non-polynomial in the field strengths $F$, $H$, $G$, and $\mathcal W$, this fact goes some way toward generating all of the higher-order terms.
Finally, the tight structure of this formulation of eleven-dimensional supergravity seems well-suited to the study of higher-derivative corrections to the action, although this may be easier to demonstrate once we have presented the non-linear couplings \cite{nonlinear}. 
(The coupling of the gravitino multiplets to conformal supergravity of \cite{Kuzenko:2017ujh} might be useful in this context.)

Besides the immediate extensions just mentioned, and which are needed to really complete the embedding of M-theory into 4D, $N=1$ superspace, there are some applications of this result and lower-dimensional analogues which could be worked out. First, closely related to the present story should be the superspace description of type IIA string theory compactified on {$G_2$}-structure manifolds. Our choice to focus on M-theory on {$G_2$}-structure backgrounds was motivated by minimality: Eleven-dimensional supergravity has the most economic field content of all higher-dimensional supergravity theories, and 4D, $N=1$ is the most familiar superspace. Describing type IIA string theory in superspace is potentially messier given the larger number of fields but can, in principle, be obtained from our formalism by dimensional reduction.

Potentially less straightforward is the description of type IIB string theory on {$G_2$}-structure backgrounds. 
It would be interesting to work this out to elucidate how mirror symmetry is realized in superspace. In fact, quite recently, a proposal for mirror symmetry for {$G_2$}-manifolds applicable to the twisted connected sum construction of ref. \cite{Corti:2012kd}
was made in ref. \cite{Braun:2017ryx}. It should then be possible to find a map between the
(super)space-time actions for type IIA and type IIB string theory compactified on mirror {$G_2$}-manifolds,
resembling the c-map in ref.\cite{Cecotti:1988qn}.  In addition to mirror symmetry, other dualities can be considered.  In the context of duality between M-theory, heterotic, and F-theory we could try to make contact with the recent paper~\cite{Braun:2017uku}, at least in the case of smooth manifolds.

Another scenario that would be worth exploring in our formalism is the case when the internal manifold has a resolved orbifold singularity, as discussed in~\cite{Acharya:2002kv}.  For a local model, we would consider internal manifolds of the form $M\times Q$, where $M$ is a resolved ADE singularity and $Q$ is a three-manifold.  Away from the singular point, the massless fields give only a $\operatorname{U}(1)^r$ gauge group corresponding to the harmonic 2-forms on $M$, but as we approach the singular limit, various massive fields become light and the gauge group enhances to something non-abelian.  In terms of the space-time effective action, there is a contribution to the superpotential in the form of a complex Chern-Simons invariant on $Q$, as explained in~\cite{Acharya:2002kv}.  Since we are keeping all KK-modes, we might be able to usefully study this limit, see more direct evidence for the enhancement, and compute the relevant superpotential terms.

In flux compactifications there is typically a warp factor multiplying the space-time part of the metric, and this has complicated the analysis of the effective theory, particularly for the purposes of constructing an $N=1$ superspace action~\cite{Shiu:2008ry}.  Our approach is applicable to those scenarios and has the potential to simplify the analysis substantially.

It would also be interesting to make contact with ref. \cite{Guio:2017zfn} where the space-time action for massless fields obtained from a compactification of eleven-dimensional supergravity on twisted connected sum {$G_2$}-manifolds was presented. In our analysis all fields (not only the massless ones) are taken into account so we could, for example, compute the gravitino mass matrix and analyze how it behaves in various limits (for example, as a function of the gluing modulus). It would, of course, be interesting to develop these ideas further and analyze the (super)space-time action for compactification of the extra twisted connected sum type of ref. \cite{Crowley:2015}, particularly since this might help elucidate the physical significance of the new homotopy invariant introduced by Crowley and Nordstr\"om in ref. \cite{Crowley:1211}.  Also quite recently, a new construction of {$G_2$}-holonomy manifolds was found in ref. \cite{Joyce:1707}, and it would be interesting to consider the corresponding (super)space-time action. In short, there has been a proliferation of new results in the mathematics literature concerning {$G_2$}-holonomy manifolds and it will be fascinating to work out the physical implications.

Finally, it would be desirable to understand the truncations of this formulation of eleven-dimensional supergravity to other dimensions.
This would make contact with (and potentially simplify) the phenomenological literature on five-dimensional supergravity \cite{Linch:2002wg, Buchbinder:2003qu, Paccetti:2004ri, Abe:2005ac,  Sakamura:2012bj, Sakamura:2013cqd} and extensions to six-dimensions\cite{Abe:2015bqa, Abe:2015yya, Abe:2017pvw}.

\section*{Acknowledgements}
We thank Andy Royston for helpful insights and many motivating discussions.
This work is partially supported by NSF under grants PHY-1521099 and PHY-1620742 and the Mitchell Institute for Fundamental Physics and Astronomy at Texas A\&M University.
We also thank the Simons Center for Geometry and Physics and the organizers of the September 2017 Workshop on Special Holonomy, where results from this work were reported.

\appendix
\section{$G_2$ in a Nutshell}
\label{S:G2}

In this appendix, we collect some useful definitions and formul\ae{} of $G_2$-structure manifolds \cite{joyce2000compact, Hitchin:2000jd, Hitchin2001, Bryant2003, Karigiannis:0301218}.
Let $\varphi$ be a 3-form on $Y$ and define the symmetric bilinear form $g_{ij}(\varphi)$ through the non-linear equation
\begin{align}
\label{E:metric}
\sqrt{g} g_{ij} :=  -\tfrac1{144} \epsilon^{abcdefg}\varphi_{iab} \varphi_{cde} \varphi_{jfg}
{~~},
\end{align}
where $g=\mathrm{det}(g_{ij})$.
The 3-form $\varphi$ is \emph{stable} if this determinant is non-zero everywhere and \emph{positive} if, in addition, $g_{ij}$ is a Riemannian metric.  These are open conditions, so if we start at a three-form for which they hold, then they will also hold for nearby three-forms.
Throughout this paper, we will assume that these conditions hold without further qualification.

Using the metric, we define the Hodge dual
\begin{align}
{\tilde \varphi} : = \ast \varphi
{~~}.
\end{align}
This equation is highly non-linear in $\varphi$ since the $\ast$ operation is non-linear in $g(\varphi)$ which is, itself, non-linear in $\varphi$.
The tensors $\varphi$, $g(\varphi)$, and ${\tilde \varphi}(\varphi)$ satisfy the algebraic identities
\begin{align}
\label{E:contractions}
&{\tilde \varphi}^{ijkl}{\tilde \varphi}_{i'j'k'l} =
	6\delta^{[i}_{[i'} \delta^{j}_{j'}\delta^{k}_{k]}
	-\varphi_{i'j'k'}\varphi^{ijk}
	-9 \delta^{[i}_{[i'}{\tilde \varphi}_{j'k']}{}^{jk]}
{~},
\cr
&\varphi^{ijk} \varphi_{ij'k'} = 2\delta_{[j'}^j\delta_{k']}^k - {\tilde \varphi}_{j'k'}{}^{jk}
{~},~~
{\tilde \varphi}^{ijkl} {\tilde \varphi}_{ijk'l'} = 8 \delta_{[k'}^k \delta_{l']}^l - 2 {\tilde \varphi}_{k'l'}{}^{kl}
{~},
\cr
&\varphi^{ijk} \varphi_{ijk'} = 6\delta_{k'}^k
{~},~~
{\tilde \varphi}^{ijkl} {\tilde \varphi}_{ijkl'} = 24 \delta_{l'}^l
{~},~~
\varphi_i{}^{lm} {\tilde \varphi}_{jk lm} = - 4 \varphi_{ijk}
{~},
\end{align}
where indices are raised and lowered with the metric (\ref{E:metric}).

A stable 3-form on the tangent space of $Y$ reduces the structure group $GL(7) \to G_2$ so that $Y$ is a $G_2$-structure manifold.
Under this reduction, the $\mathbf{21}$-dimensional space of 2-forms on $Y$ decomposes into $G_2$ representations as $\mathbf{21}=\mathbf{7}\oplus \mathbf{14}$.
Similarly, the $\mathbf{35}$-dimensional space of 3-forms on $Y$ decomposes as $\mathbf{35}=\mathbf{1}\oplus\mathbf{7}\oplus \mathbf{27}$.
For any $p$-form $\omega$, let $\omega_\mathbf i := \pi_{\bm {i}} \omega$ denote the projection to the $\mathbf{i}$-dimensional representation.
Explicitly, for any 2-form $\eta$ and 3-form $\omega$,
\begin{subequations}\label{E:G2Projs}
\begin{align}
\pi_{\bm 7} \eta_{ij} &= \left( \tfrac13\delta_i^k \delta_j^l  - \tfrac16{\tilde \varphi}_{ij}{}^{kl} \right) \eta_{kl}
{~},\\
\pi_{\bm {14}} \eta_{ij} &= \left( \tfrac23\delta_i^k \delta_j^l  +\tfrac16 {\tilde \varphi}_{ij}{}^{kl} \right) \eta_{kl}
{~},\\
\label{E:35to1}
\pi_{\bm {1}} \omega_{ijk} &= \tfrac1{42} \varphi_{ijk} \varphi^{i'j'k'} \omega_{i'j'k'}
{~},\\
\label{E:35to7}
\pi_{\bm {7}} \omega_{ijk} &=  \left(
	\tfrac14\delta_i^{i'}\delta_j^{j'}\delta_k^{k'}
	-\tfrac38 {\tilde \varphi}_{[ij}{}^{i'j'}\delta_{k]}^{k'}
	-\tfrac1{24}\varphi_{ijk} \varphi^{i'j'k'}\right) \omega_{i'j'k'}
{~},\\
\pi_{\bm {27}} \omega_{ijk} &=  \left(
	\tfrac34\delta_i^{i'}\delta_j^{j'}\delta_k^{k'}
	+\tfrac38 {\tilde \varphi}_{[ij}{}^{i'j'}\delta_{k]}^{k'}
	+\tfrac1{56}\varphi_{ijk} \varphi^{i'j'k'}\right) \omega_{i'j'k'}
{~~}.	
\end{align}
\end{subequations}
The $\bm 7$-projections of 2- and 3-forms play an important role in the gravitino analysis. We define for such projections the vectors fields\footnote{That is, for any $\eta \in \Lambda^2(Y)$ and $\omega \in \Lambda^3(Y)$, we are defining the vectors $\vec{\eta}$ and $\vec{\omega}$ on $Y$ such that $\iota_{\vec{\eta}} \varphi = \pi_{\bm 7} \eta$ and
$\iota_{\vec{\omega}} {\tilde \varphi} = 2 \pi_{\bm 7} \omega$.} 
\begin{subequations}
\begin{align}
\label{E:7components2}
\eta^i:=\tfrac16 \varphi^{ijk} \pi_{\bm 7} \eta_{jk} = \tfrac16 \varphi^{ijk} \eta_{jk}
~~~&\Leftrightarrow~~~ \pi_{\bm 7} \eta_{ij} = \varphi_{ijk} \eta^k
\\
\label{E:7components3}
\omega^i := \tfrac1{12} {\tilde \varphi}^{ijkl} \pi_{\bm 7} \omega_{jkl}= \tfrac1{12}{\tilde \varphi}^{ijkl} \omega_{jkl}
~~~&\Leftrightarrow~~~
\pi_{\bm 7} \omega_{ijk} = -\tfrac12 {\tilde \varphi}_{ijkl} \omega^l
{~~}.
\end{align}
\end{subequations}
Note that this implies that there are conversion factors in squares
\begin{align}
\label{E:squares}
(\eta^i)^2  = \tfrac16  (\pi_{\bm 7} \eta_{ij})^2
~~~\textrm{and}~~~
(\omega^i)^2  = \tfrac16 (\pi_{\bm 7} \omega_{ijk})^2
{~~}.
\end{align}
These factors appear when we confirm the coefficients of the kinetic terms of all gauge fields in the $\bm 7$ in (\ref{E:7Kinetic}).
The dual 4-form $\tilde \varphi$ acts on 2-forms as $\tilde \varphi_{ij}{}^{kl} \eta_{\bm 7 kl} = -4 \eta_{\bm 7 ij}$ and $\tilde \varphi_{ij}{}^{kl} \eta_{\bm {14} kl} = 2\eta_{\bm {14} ij}$
or 
\begin{align}
\label{E:2formSquare}
\tilde \varphi^{ij kl} \eta_{ij} \eta_{kl} 
= -4 \eta_{\bm 7 ij}^2 + 2 \eta_{\bm {14} ij}^2
= -24 (\eta^i)^2 + 2 \eta_{\bm {14} ij}^2
{~~}.
\end{align}
Momentarily, we will use similar equations on the space of 3-forms,
\begin{align}
\label{E:3formSquares}
\omega^{ijk}\omega_{ijk} &=
	\omega_\mathbf 1^2 + \omega_\mathbf 7^2 + \omega_{\mathbf {27}}^2
\cr
g^{ii'} {\tilde \varphi}^{jk j'k'} \omega_{ijk}\omega_{i'j'k'} &=
	- 4 \omega_\mathbf 1^2 -2 \omega_\mathbf 7^2 + \tfrac23 \omega_{\mathbf {27}}^2
\cr
(\varphi^{ijk}\omega_{ijk})^2 &=
	42 \omega_\mathbf 1^2
{~~}.
\end{align}

The Hitchin functional is defined as the Riemannian volume
\begin{align}
S_H = \int d^7 y\, \sqrt{g(\varphi)}
{~~}.
\end{align}
Since this will be the main ingredient in our K\"ahler term, it will prove useful to derive the first few functional derivatives.
The first derivative is the dual of $\varphi$
\begin{align}
3! \epsilon_{ijklmnp} {\delta K\over \delta F_{mnp} }
	= \tilde \varphi_{ijkl}
{~~},
\end{align}
and the second derivative is (proportional to) the Hitchin metric on the moduli space of $G_2$ structures \cite{Hitchin:2000jd} 
\begin{align}
\label{E:HitchinMetric}
G^{ijk,mnp} &:= - \frac{\partial^2 \sqrt{g(\varphi)}}{\partial \varphi_{ijk}\partial \varphi_{mnp}}
= \tfrac1{3!\cdot 3! } \sqrt{g} \left(
	g^{[i|m}g^{|j|n}g^{|k]p}
 	+\tfrac1{18}{\varphi}^{ijk}{\varphi}^{mnp}
	+\tfrac3{2} g^{[m|[i}{\tilde \varphi}^{jk]|np]}
	\right)
{~}.	
\end{align}
The contractions (\ref{E:3formSquares}) can be used to compute the signature
\begin{align}
\label{E:3formMetric}
18 \omega_{ijk} G^{ijk,lmn} \omega_{lmn} &=
	-\tfrac43 \omega_{\mathbf 1 ijk}^2
	-\omega_{\mathbf 7 ijk}^2
	+\omega_{\mathbf{27} ijk}^2
\end{align}
in terms of $G_2$ projections for any 3-form $\omega$.

It will also be useful to introduce the intrinsic torsion forms $\tau_\mu$ for $\mu=0,1,2,3$ and analogous quantities $\sigma_\mu$ for $\mu=0,1,3$ defined by \cite{Bryant2003}
\begin{subequations}
\begin{align}
\label{E:TorsionClassesDef}
d\varphi &= \tau_0 {\tilde \varphi} + 3\tau_1\varphi + \ast \tau_3
{~},~~
d{\tilde \varphi} = 4 \tau_1 {\tilde \varphi} + \tau_2 \varphi
{~},
\\
dC &= \sigma_0 {\tilde \varphi} + 3\sigma_1\varphi + \ast \sigma_3
{~},
\end{align}
\end{subequations}
where the subscripts indicate the degree as a form, and where we impose that $\tau_2$ transforms in the $\mathbf{14}$ (so $\pi_{\mathbf{14}}\tau_2=\tau_2$) and that $\tau_3$ and $\sigma_3$ transform in the $\mathbf{27}$.
(We could make the analogous definition for the the components of $d \ast C$ but the action depends only on $C$ and $dC$; the $C$-field analogue of the torsion class $\tau_2$ is not gauge invariant.)

\section{Old Minimal Supergravity}
\label{S:OMSG}

In this section we review the elements of old minimal supergravity (see e.g.
\cite{Gates:1983nr, Wess:1992cp,Buchbinder:1998qv})
used in this work.
The component fields of spin 2 and $\tfrac 32$ are described in terms of the superspace analog of the conformal Weyl tensor $C_{abcd}$ (the trace-free part of the Riemann tensor $R_{abcd}$).
Converting to spinor indices, $C_{\alpha \beta \gamma \delta}=C_{(\alpha \beta \gamma \delta)}$ is totally symmetric.
(Any anti-symmetric part can be isolated with $\varepsilon_{\alpha \beta}$ and corresponds to a trace.)
Its spin-$\tfrac32$ analog is the gravitino curl $W_{\alpha \beta \gamma}=W_{(\alpha \beta \gamma)}$.
Together, they are contained within a superfield $W_{\alpha \beta \gamma}$ (for which the gravitino curl is the bottom component) subject to the conditions
\begin{align}
\bar D_{\dt \alpha} W_{\alpha \beta \gamma}=0
~~~\textrm{and}~~~
\partial_{\dt\alpha}{}^{\beta} D^{\gamma} W_{\gamma\beta\alpha}
= -\partial_{\alpha}{}^{\dt\beta} \bar D^{\dt\gamma} \bar W_{\dt\gamma\dt\beta\dt\alpha}~.
\end{align}
Together these imply that $W_{\alpha \beta \gamma}$ contains, in addition to the gravitino field strength, the component Weyl tensor as $C_{\alpha \beta \gamma \delta}=D_{(\delta} W_{\alpha \beta \gamma)}$ and a $U(1)_R$ field strength as $D^\gamma W_{\gamma\beta\alpha}$. For old minimal supergravity, the $U(1)_R$ connection is auxiliary and pure gauge, so that
\begin{align}
D^\gamma W_{\alpha \beta \gamma} \stackrel{eom}= 0{~},
\end{align}
where $\stackrel{eom}= 0$ indicates that this combination vanishes only on-shell.

The full set of Bianchi identities is an off-shell version of these constraints.
They may be solved in terms of the conformal supergravity prepotential $H^a$.
(We will need only the linearized expressions.)
Converting its 4-vector index into a bi-spinor index using the Pauli matrices, the conformal supertensor is given in terms of it as
\begin{align}
\label{E:GravitonRep}
W_{\alpha \beta \gamma}= \tfrac i{8}\bar D^2 D_{(\alpha}\partial_\beta{}^{\dt \gamma} H_{\gamma)\dt \gamma}
~~~\Rightarrow~~~
\delta H_{\gamma \dt \gamma} = \bar D_{\dt \gamma} L_\gamma - D_\gamma \bar L_{\dt \gamma}
{~~}.
\end{align}
This combination of $D$'s projects onto the desired irreducible superspin-$\tfrac32$ representation consisting of component spins $(\tfrac32, 2)$ \cite{Siegel:1981ec}.
Equivalently, it is invariant under the huge gauge transformation involving the unconstrained superfield parameter $L_\alpha$, which allows one to gauge away all but 
the spin-2 conformal graviton and spin-$\tfrac32$ conformal gravitino.

For Poincar\'e supergravity, this gauge transformation is too large:
We are required to reinstate the spin-0 part of the graviton (trace) and the spin-$\tfrac12$ part of the gravitino (gamma-trace).
A closely-related statement is that it is not possible to construct a two-derivative action from this representation alone.\footnote{The conformal supergravity action $\int d^2\theta \,W^{\alpha \beta \gamma} W_{\alpha \beta \gamma}$ is the supersymmetrization of the four-derivative Weyl$^2$ action.}
Following \cite{Siegel:1978mj}, this is done by coupling conformally to a superfield with a scalar component that has a non-vanishing background value.
This scalar field is the conformal compensator.
Different off-shell supergravity theories correspond to different choices for this scale compensator.\footnote{For a particularly enlightening classification of irreducible quadratic Poincar\'e supergravity actions and treatment of scale compensators, see \cite{Gates:2003cz}.}
Old minimal supergravity involves a chiral scalar superfield $\Phi_0=e^\sigma$, often written in an exponential form as its background value is taken to be 1. Its linearized gauge transformation is
\begin{align}
\delta \sigma = -\frac 1{12} \bar D^2 D^\alpha L_\alpha
{~~}.
\end{align}
The quadratic action of old minimal supergravity is \cite{Gates:1983nr, Buchbinder:1998qv}
\begin{align}
\label{E:OMSG}
S_{OMSG} &= \frac1{\kappa^2} \int d^4 x \int d^4\theta \, L_{OMSG}{~}, \\
L_{OMSG} &= \frac18 H_a D^\beta \bar D^2 D_{\beta} H^a
	- (\partial_a H^a)^2
	+ \frac1{48}([D_\alpha, \bar D_{\dt \alpha}] H^{\alpha \dt \alpha})^2
	-3 \bar \sigma \sigma
	+2i (\sigma-\bar \sigma)\partial_a H^a
~.
\nonumber
\end{align}

There is a modification \cite{Grisaru:1981xm, Gates:1980az} of this action in which the conformal chiral compensator $\Phi_0$ is replaced with a slightly different representation in terms of a real prepotential. 
Generally, the chirality constraint on a generic chiral field can be solved in terms of a complex scalar superfield $X_{\mathbf C}$,
\begin{align}
\bar D_{\dt \alpha} \Phi = 0
~~~\Rightarrow~~~
\Phi = -\tfrac1{4} \bar D^2 X_{\mathbf C}~.
\end{align}
But a closer inspection of the components reveals that the same {\em physical} components result from the restriction $\bar X_{\mathbf C} = X_{\mathbf C}=: X$.
In this representation, the scalar that was the imaginary part of the F-component of $\Phi$ is replaced by the divergence of a vector,
\begin{align}
i D^2 \Phi - i \bar D^2 \bar \Phi = -\tfrac i4 [D^2, \bar D^2] X
	= - \partial^{\alpha \dt \alpha} ( [D_\alpha , \bar D_{\dt \alpha}] X)
{~},
\end{align}
or, equivalently, the dual of a four-form field strength.
This is just the superspace representation of the gauge 3-form $C_{abc}$ \cite{Gates:1980ay}, and one recognizes the superfield $G$ and its gauge 3-form prepotential $X$.
It was already observed in \cite{Becker:2016edk} that this field strength plays the role of the conformal compensator (at least in the gauge where $H_i \to 0$).
This suggests that the 4D, $N=1$ supergravity formulation best suited to the description of eleven-dimensional supergravity is given by this modification of old minimal supergravity \cite{Grisaru:1981xm, Gates:1980az}  with the replacements\footnote{The relative normalizations of $\sigma$ and $G$ arise because $G$ has conformal weight 3 while $\Phi_0$ is normalized to have weight 1.}
\begin{align}
e^{3 \sigma} \to  G = -\tfrac14 \bar D^2 X
~~~\textrm{with}~~~
\delta_L X = D^\alpha L_\alpha + \bar D_{\dt \alpha}\bar L^{\dt \alpha}
{~~}.
\end{align}

The $L_\alpha$ gauge invariance can be exploited in several different ways.
One choice is to fix $\sigma\to 0$. Another choice is to impose the Lorentz gauge
\begin{align}
\label{E:LorentzGauge}
D^\alpha H_{\alpha \dt \alpha} \to 0
~~~\Rightarrow~~~
L_{OMSG} \to - H_a \Box H^a
	-3 \bar \sigma \sigma
{~~}.
\end{align}
In this gauge, the component action reduces to that of linearized supergravity in transverse-traceless gauge. Both of these are manifestly supersymmetric gauge choices, but leave some unphysical component fields unfixed.

\subsection{Wess-Zumino Gauge and Components}
\label{S:SGWZ}
It is usually more convenient to impose a Wess-Zumino gauge choice that eliminates all but the physical components. Using a vertical bar $\vert$ to denote projection to $\theta=0$, fixing certain components of $H_a$ to zero restricts the superfield $L_\alpha$:
\begin{subequations}
\begin{align}
H_{\alpha {\dt\alpha}}\vert &= 0 \quad \Rightarrow \quad
	D_\alpha \bar L_{\dt\alpha}\vert = \bar D_{\dt\alpha} L_\alpha\vert{~}, \\
D^{\beta} H_{\beta {\dt\alpha}}\vert &= 0 \quad \Rightarrow \quad
	D^2 L_{\dt\alpha}\vert = -\bar D_{\dt\alpha} D^\alpha L_\alpha\vert{~},\\
D_{(\beta} H_{\alpha) {\dt\alpha}}\vert &= 0 \quad \Rightarrow \quad
	D_{(\beta} \bar D_{\dt\alpha} L_{\alpha)} \vert = \bar D_{\dt\alpha} D_{(\beta} L_{\alpha)} = 0{~}, \\
D^2 H_{\alpha \dt\alpha}\vert &= 0 \quad \Rightarrow \quad D^2 \bar D_{\dt\alpha} L_\alpha = 0~.
\end{align}
\end{subequations}
In addition, the ability to shift $L_\alpha$ by a chiral spinor implies we can always take $L_\alpha\vert = D_\beta L_\alpha\vert = D^2 L_\alpha \vert = 0$.
At the $\theta \bar\theta$ level, we identify the spin-2 part of the graviton as
\begin{align}
\label{E:ConfMetric}
h^{\text{spin-2}}_{\beta \dt\beta\, \alpha \dt\alpha \vphantom{\dot\beta}}
	:= -[D_{\beta}, \bar D_{\dt\beta}] H_{\alpha \dt\alpha} \vert
\end{align}
and impose another WZ condition to ensure that the right-hand side is symmetric
in $\beta\alpha$ and $\dt\beta\dt\alpha$ so that $h^{\text{spin-2}}_{ba}$ is symmetric
and traceless. This fixes
\begin{align}
D_{(\beta} \bar D^2 L_{\alpha)} \vert = 0{~}, \qquad
\text{Re } D^\alpha \bar D^2 L_\alpha \vert = \partial_a \xi^a{~},
\end{align}
where $\xi^m$ is the linearized diffeomorphism
\begin{align}
\xi_{\alpha {\dt\alpha}} := -i (\bar D_{\dt\alpha} L_\alpha + D_\alpha \bar L_{\dt\alpha}) \vert{~}, \qquad
\delta h^{\text{spin-2}}_{a b} = 2 \partial_{(a} \xi_{b)} - \frac{1}{2} \eta_{a b} \partial_c \xi^c~.
\end{align}
The $N=1$ supersymmetry parameter is 
\begin{align}
\xi_\alpha = -\frac{1}{4} \bar D^2 L_\alpha\vert = \frac{1}{4} D_\alpha \bar D_{\dt\alpha} L^{\dt\alpha} \vert{~},
\end{align}
and the spin-$\tfrac32$ part of the $N=1$ gravitino is
\begin{align}
\psi^{\text{spin-$\tfrac32$}}_{\beta\dt\beta\, \alpha \vphantom{\dot\beta}}
	= -\frac{i}{4} \bar D^2 D_{(\beta} H_{\alpha) \dt\beta} \vert{~}, \qquad
\delta \psi_{\beta \dt\beta\, \alpha} = 2 \partial_{\dt\beta (\beta} \xi_{\alpha)}~.
\end{align}
The final WZ condition guarantees that no other fermions appear at the $\theta \bar\theta^2$ level,
\begin{align}
\bar D^2 D^\alpha H_{\alpha \dt\alpha}\vert = 0 \quad \Rightarrow \quad
\bar D^2 D^2 \bar L_{\dt\alpha} \vert = 2 i \partial_{\alpha \dt\alpha} \bar D^2 L^\alpha\vert~.
\end{align}
The top component of $H_{\alpha\dt\alpha}$ corresponds to the $U(1)_R$ gauge field,
\begin{align}
d_{\alpha\dt\alpha} = -\frac{1}{16} D^\beta \bar D^2 D_\beta H_{\alpha\dt\alpha}\vert{~}, \qquad
\delta d_a = \partial_a \omega{~}, \qquad
\omega = -\frac{1}{8} \,\text{Im } D^\alpha \bar D^2 L_\alpha \vert~.
\end{align}

While $H_{\alpha\dt\alpha}$ contains the component field content of $N\!=\!1$ conformal supergravity, the compensator $G$  contains a propagating complex scalar and Weyl fermion. These transform as
\begin{align}
\delta G \vert &= -\frac{1}{4} \bar D^2 D^\alpha L_\alpha\vert
	= \frac{3}{4} \partial_a \xi^a + 2 i \omega{~}, \\
\delta\, D_\alpha G \vert &=  -\frac{1}{4} D_\alpha \bar D^2 D^\beta L_\beta\vert
	= 3 i \, \partial_{\alpha \dt\beta} \bar\xi^{\dt\beta}~.
\end{align}
We identify the spin-0 part of the metric as the real part of $G\vert$, and the spin-$\tfrac12$ part of the gravitino as the fermion $D_\alpha G\vert$, so that
\begin{subequations}
\begin{align}
\label{E:Metric}
h_{b a}
	&:= \frac{1}{2} \sigma_b^{\dt\beta \beta} [D_{\beta}, \bar D_{\dt\beta}] H_{a} \vert
	+ \frac{1}{3} \eta_{b a} (G + \bar G) \vert{~}, \\
\label{E:Gravitino}
\psi_{\beta \dt\beta\, \alpha} &:= -\frac{i}{4} \bar D^2 D_{(\beta} H_{\alpha) \dt\beta} \vert
	- \frac{i}{3} \epsilon_{\beta \alpha} \bar D_{\dt\beta} \bar G \vert~.
\end{align}
The $U(1)_R$ gauge transformation associated with $\omega$ may be used to set the imaginary part of $G\vert$ to zero; equivalently, it is eaten by the auxiliary gauge field $d_a$.\footnote{An alternative WZ gauge-fixing involves setting $G\vert = 0$ and $D_\alpha G\vert = 0$. Then the spin-0 part of the metric and spin-$\tfrac12$ part of the gravitino are also contained within $H_a$. This alters the definitions \eqref{E:Metric}, \eqref{E:Gravitino} but leaves the Lagrangian \eqref{E:OMSGComponents} unchanged.}
The remaining degrees of freedom are the $\theta^2$ components of $G$. These
are the real auxiliary $d_X$ and the 4-form field strength $F_{abcd}$, given by
\begin{align}
F_{abcd} &:= \tfrac i8 \epsilon_{abcd} (D^2 G - \bar D^2 \bar G)\vert
\\
d_X &:= \tfrac1{32} \{ D^2 , \bar D^2 \} X\vert
{~~}.
\end{align}
\end{subequations}
Then the component Lagrangian for modified old minimal supergravity can be written
\begin{align}
\label{E:OMSGComponents}
\kappa^2 L_{OMSG}=
	-\tfrac18 (\partial_c h_{ab})^2
	+ \tfrac14 (\partial^b h_{ab})^2
	+ \tfrac18 (\partial_a h)^2
	+ \tfrac14 h \partial^a \partial^b h_{ab}
\cr
	+\tfrac43 d_a^2
	-\tfrac13 \left( d_X^2 + F_{abcd}^2 \right)
{~~}.
\end{align}
We will use this when comparing to the linearized eleven-dimensional action in \S{}\ref{S:Components}.

\section{Gravitino Superfields}
\label{S:4DGravitino}

In this section, we will work out in detail the quadratic superspace actions for a free spin-$\tfrac32$ field and its spin-1 superpartner.
In manifestly supersymmetric language, we are considering the free massless superspin-1 multiplet which is often referred to as a (matter) gravitino multiplet.
Investigations into the precise form of this action date back to the work of Ogievetsky and Sokatchev \cite{Ogievetsky:1975vk}.
A second formulation was discovered by de Wit and van Holten \cite{deWit:1979pq} and Fradkin and Vasiliev \cite{Fradkin:1979as}.
The relation between these theories was elucidated at the level of superfield representation theory in \cite{Gates:1979gv}, at the level of supergeometry in \cite{Gates:1984mt}, and from 4D, $N=2$ superspace in \cite{Butter:2010sc}. 
In \cite{Linch:2002wg}, the 5D gravitino superfield was discovered to be neither of these multiplets.

In modern terms, the basic matter gravitino model can be defined by the conformal gravitino field strength $W_{\alpha \beta i}$ describing a set of irreducible super-spin-1 superfields, with $i$ denoting the additional gravitini. On-shell such a representation consists of a spin-$\tfrac32$ gravitino and a spin-1 ``graviphoton''. 
This implies that the superfield satisfies
\begin{align}
\bar {D}_{\dt \alpha} W_{\alpha \beta i} = 0
~~~\textrm{and}~~~
{D}^\beta W_{\alpha \beta i} \stackrel{eom}= 0
{~~},
\end{align}
where the second equation is required to hold only on-shell.\footnote{There is 
a dual formulation of the superspin-1 multiplet with prepotential $\Psi_{\alpha \beta \dt \beta}$ and
field strength $W_{\alpha \beta} \sim \bar D^{\dt \gamma} D_{(\alpha} \left[ 2D^\gamma \Psi_{\beta)\gamma \dt \gamma} - \bar D^{\dt \beta} \bar \Psi_{\beta) \dt \beta \dt \gamma}\right]$ \cite{Kuzenko:1993jq} (see also \S{}6.9 of \cite{Buchbinder:1998qv}).
We thank Sergei Kuzenko for emphasizing this point to us.
} 
We suppress the full set of off-shell Bianchi identities which can be solved in terms of an unconstrained gravitino prepotential superfield as
\begin{align}
\label{E:GravitinoFS}
W_{\alpha \beta i} = -\tfrac14 \bar {D}^2 {D}_{(\alpha}\Psi_{\beta) i}
{~~}.
\end{align}
This expression has a large (pre-)gauge symmetry
\begin{align}
\label{E:GravitinoTransformations}
\delta \Psi_{\alpha i} = \Xi_{\alpha i} + {D}_\alpha \Omega_i
~~~\textrm{with}~~~
\bar {D}_{\dt \alpha} \Xi_{\alpha i} = 0
\end{align}
and $\Omega_i$ complex and unconstrained. We will be interested in the case where $\Psi_{\alpha i}$ transforms also under the $N=1$ conformal supergravity $L_\alpha$ transformation as in \eqref{E:GravitinoXf}. Before discussing the actions and compensating mechanisms in detail, it is useful to discuss the off-shell components of $\Psi_{\alpha i}$.

\subsection{Wess-Zumino analysis}
\label{S:WZGravitino}
The gravitino superfield $\Psi_{\alpha i}$, like the $N=1$ supergravity prepotential $H_a$, is subject to a large set of gauge transformations, here encoded in the parameters $\Xi_{\alpha i}$ and $\Omega$. The gauge parameter $\Omega$ appears under $D_\alpha$, which means that it is defined only up to a shift by an antichiral superfield. This means we may assume (without loss of generality) $\Omega_i \vert = \bar D_{\dt\alpha} \Omega_i \vert = \bar D^2\Omega_i \vert = 0$. Turning to $\Psi_{\alpha i}$ itself, a number of WZ conditions can be imposed. These in turn constrain the residual gauge symmetries within $\Xi$ and $\Omega$:
\begin{subequations}
\begin{align}
\Psi_{\alpha i}\vert  &= 0 \quad \Rightarrow \quad \Xi_{\alpha i} \vert = -D_\alpha \Omega_i \vert {~}, \\
D^\alpha \Psi_{\alpha i}\vert  &= 0 \quad \Rightarrow \quad D^\alpha \Xi_{\alpha i} \vert = -D^2 \Omega_i \vert {~}, \\
D_{(\beta} \Psi_{\alpha) i}\vert  &= 0 \quad \Rightarrow \quad D_{(\beta} \Xi_{\alpha) i} \vert = 0{~}, \\
\bar D_{\dt\alpha} \Psi_{\alpha i} \vert&= 0 \quad \Rightarrow \quad
	\bar D_{\dt\alpha} D_\alpha \Omega_i \vert 
	= \partial_i \xi_{\alpha \dt\alpha}{~}, \\
D^2 \Psi_{\alpha i} \vert&= 0 \quad \Rightarrow \quad D^2 \Xi_{\alpha i} \vert = 0{~}, \\
D^\alpha \bar D_{\dt\alpha} \Psi_{\alpha i} \vert&= 0 \quad \Rightarrow \quad
	D^\alpha \bar D_{\dt\alpha} D_\alpha \Omega_i \vert
	= 8 i \,\partial_i \bar\xi_{\dt\alpha}{~}, \\
\bar D^2 \Psi_{\alpha i} \vert &= 0 \quad \Rightarrow \quad
	\bar D^2 D_\alpha \Omega_i \vert = 8i \,\partial_i \xi_\alpha{~}, \\
\bar D^2 D^\alpha \Psi_{\alpha i} \vert &= 0 \quad \Rightarrow \quad
	\bar D^2 D^2 \Omega_i \vert =  6 i \, \partial_i \partial_a \xi^a~.
\end{align}
\end{subequations}
Note that certain components of $\Omega_i$ are related to the internal derivatives of the diffeomorphism and $N=1$ supersymmetry parameters $\xi_a$ and $\xi_\alpha$. This is a consequence of the $\partial_i L_\alpha$ term in the gravitino superfield transformation (\ref{E:GravitinoXf}).

The spin-$\tfrac32$ part of the extended gravitino is defined as the remaining $\theta\bar\theta$
component:
\begin{align}
\label{E:ExtSUSY}
\psi_{(\beta \dt\beta \,\alpha) i} := -i D_{(\alpha} \bar D_{\dt\beta} \Psi_{\beta) i}\vert{~},\qquad
\delta \psi_{(\beta \dt\beta \,\alpha) i} = 2 \,\partial_{\dt\beta (\beta} \xi_{\alpha) i}\vert{~}, \qquad
\xi_{\alpha i} := \Xi_{\alpha i} \vert~.
\end{align}
The remaining components of $\Psi_{\alpha i}$ are auxiliary fields that can be defined as
\begin{subequations}
\begin{align}
\label{E:Auxy}
y_{\alpha \dt\alpha \,i} &:= -\tfrac14 {D}^2 \bar D_{\dt\alpha} \Psi_{\alpha i}\vert{~}, \\
t_{\alpha \beta i} &:= -\frac{1}{4} \bar D^2 D_{(\alpha} \Psi_{\beta) i}\vert{~}, \\
\rho_{\alpha i} &:= \frac{1}{3} D^\beta \bar D^2 D_{(\alpha} \Psi_{\beta) i}\vert~.
\end{align}
\end{subequations}
The component $t_{\alpha \beta i} = W_{\alpha\beta i}\vert$ describes an anti-self-dual rank-two tensor, and $\rho_{\alpha i}$ must play the role of a Lagrange multiplier due to its high dimension. Being $\Omega$ and $\Xi$ invariant, these only transform under the $L_\alpha$ transformations. The auxiliary vector $y_{a i}$ is in contrast subject to complex gauge transformations
\begin{align}
\delta y_{a i} = \frac{i}{2} \partial_a D^2 \Omega_i \vert~.
\end{align}
In Wess-Zumino gauge, the only residual gauge symmetries are the extended supersymmetry (\ref{E:ExtSUSY}) and the bosonic symmetry associated with $D^2 \Omega_i \vert$.

The residual gauge transformation of the gravitino multiplet associated with the 7 complex parameters $D^2 \Omega_i\vert$ will allow 14 of the residual bosonic fields to be eliminated, or equivalently, eaten by the auxiliary field $y_{a i}$. We have already mentioned that the bottom component of $H_i$ must be unphysical. Its gauge transformation turns out to be
\begin{align}
\delta H_i\vert
	&= \frac{i}{2} D^\alpha \Xi_{\alpha i}\vert
	- \frac{i}{2} \bar D_{\dt\alpha} \bar\Xi_i^{\dt\alpha}\vert
	= \text{Im} D^2\Omega_i\vert{~},
\end{align}
and so one can exploit half of the residual gauge symmetry of the gravitino multiplet to eliminate it. This fixes $D^2\Omega^i\vert = \bar D^2\bar\Omega^i\vert$. The other scalar fields are contained within $\Phi_{ijk}\vert$. These transform as
\eqref{E:Grinolinb}, which in the gauge $H_i\vert=0$ implies
\begin{align}
\delta \Phi_{i j k}\vert
	&= -\frac{i}{2} \tilde\varphi_{ijkl} \,\text{Re}\,D^2 \Omega^l\vert~.
\end{align}
This ensures that only the imaginary part of $\Phi_{ijk}\vert$ transforms,
\begin{align}
\label{E:NoGrav7}
\delta F_{ijk} &= -\frac{1}{2} \tilde\varphi_{ijkl} \,\text{Re}\,D^2 \Omega^l\vert
\end{align}
while the 3-form $C_{ijk}$ is invariant. The residual gauge symmetry associated with $\text{Re}\,D^2 \Omega^l\vert$ ensures that we can eliminate the $\bm{7}$-component of $F_{ijk}$.

Finally, we find that the KK vector field transforms under $\Omega_i$ as
\begin{align}
\delta \cA^i_{\alpha{\dt\alpha}} &= \frac{1}{2} [D_\alpha, \bar D_{\dt\alpha}] \delta \cV^i\vert
	= -\frac{1}{4} [D_\alpha, \bar D_{\dt\alpha}] (\Omega^i + \bar \Omega^i)\vert
	= \partial^i \xi_{\alpha{\dt\alpha}}{~},
\end{align}
consistent with its interpretation as the component $g_{m i}$ of the 11D metric. Note this result requires a precise interplay between the $\Omega_i$ and $L_\alpha$ transformations of the gravitino superfield in WZ gauge. In contrast, the $\Omega_i$ transformation leaves the vector fields $A_{i j a}$ inert, as expected at the linearized level.

\subsection{Actions and Compensators}

Similarly to conformal supergravity, it is not possible to write a 1-derivative Rarita-Schwinger action for the conformal gravitino alone. (The only conformal invariant is the 2-derivative action $\int d^4x\, d^2\theta \,W^{\alpha \beta i} W_{\alpha \beta i}$.) To write an action, we require the analog of a scale compensator. Recalling the $N=1$ old minimal supergravity action, we would expect such a compensator to provide (in WZ gauge) the missing spin-$\tfrac12$ component of the gravitino as well as the longitudinal mode of the auxiliary vector $y_{a i}$ (\ref{E:Auxy}).

In this section we will include all possible compensator couplings from the outset. This includes superspins $\tfrac12^+\oplus\tfrac12^+\oplus\tfrac12^-\oplus0$ corresponding to real and imaginary vector multiplets, a tensor multiplet, and a scalar multiplet \cite{Gates:1979gv}. For simplicity, we will ignore any internal $y$ derivatives here. The most general quadratic action is of the form
\begin{align}
L = \int d^4 \theta \, L_D
	+ \int d^2 \theta \, L_F
	+\int d^2 \bar\theta \, \bar L_F
\end{align}
with (suppressing the index $i$ now)
\begin{subequations}
\begin{align}
\label{E:Dterm}
 L_D&=
	a_0 \bar E^{\alpha \dt \alpha} E_{\alpha \dt \alpha}
	+ a_1 E^{\alpha \dt \alpha}E_{\alpha \dt \alpha}
	+ {\bar a_1} \bar E^{\alpha \dt \alpha}\bar E_{\alpha \dt \alpha}
	+ a_2 \bar BB + a_3B^2 +\bar {a_3} \bar B^2
\cr	&+ \left[ \Psi^\alpha \left(
	a_4 \mathcal W_\alpha
	+a_5 W_\alpha
	+a_6 {D}_\alpha F
	+a_7 {D}_\alpha H\right)
	+\textrm{h.c.} \right]
	+ a_8 F^2
	+ a_9 H^2
\\
\label{E:Fterm}
L_F &= a_{10} \mathcal W^\alpha \mathcal W_\alpha
	+a_{11} W^\alpha W_\alpha
{~~}.
\end{align}
\end{subequations}
Here we have defined the complex potentials
\begin{align}
E_{\alpha \dt \alpha} := \bar {D}_{\dt \alpha} \Psi_\alpha
~~~\mathrm{and}~~~
B:= {D}^\alpha \Psi_\alpha
{~~}.
\end{align}

This action is required to be invariant under the gauge transformations (\ref{E:GravitinoTransformations}) and
\begin{align}
\delta \Sigma_\alpha = -\Xi_\alpha
{~},~~~
\delta V = \textrm{Im}\,\Omega
{~},~~~
\delta \Phi = -i \bar {D}^2 \bar \Omega
{~},~~~
\delta \mathcal V =  -\textrm{Re}\, \Omega
\end{align}
or, for the compensator ``field strengths''
\begin{subequations}
\begin{align}
\delta H &= \tfrac1{2i} \left( \bar {D} \bar \Xi  - {D} \Xi \right)
\\
\delta W_\alpha &= -\tfrac14 \bar {D}^2 {D}_\alpha \textrm{Im}\,\Omega
\\
\delta F &= - \tfrac12 \left( {D}^2 \Omega + \bar {D}^2 \bar \Omega \right)
\\
\delta \mathcal W_\alpha &= \tfrac14 \bar {D}^2 {D}_\alpha \textrm{Re}\,\Omega
{~~}.
\end{align}
\end{subequations}
The $\Xi$ transformation imposes the strong condition that $\Psi$ and $\Sigma$ appear only in the combination $\bm \Psi := \Psi + \Sigma$.
The form of the field strength $H$ then implies $(a_2, a_3, a_7,a_9) = (\tfrac c2, -\tfrac c4, ic, c )$ with $c\in \mathbf R$, so that these terms must all appear in the combination
\begin{align}
\bm H = \tfrac1{2i} \left( {D} \bm \Psi - \bar {D} \bar {\bm \Psi}\right)
	=H + \tfrac1{2i} \left( B - \bar B\right)
~~~\Rightarrow~~~	
\delta \bm H = \tfrac1{2i} \left( {D}^2 \Omega - \bar {D}^2 \bar \Omega\right)
\end{align}
as a square.
Imposing $\Omega$ invariance, we find that $(a_0, a_1)=(a,b)\in \mathbf R \times \mathbf C$ remain undetermined with $(a_4,a_5,a_6, a_8, a_{10}, a_{11}) = (2(a+b), 2i(a-b), -(a+c), a+c , a+b, a-b )$.
Then, the action can be written as
\begin{align}
\label{E:GravitinoLag}
 L_D&=
	a \bar E^{\alpha \dt \alpha} E_{\alpha \dt \alpha}
	+ \tfrac b4 E^{\alpha \dt \alpha}E_{\alpha \dt \alpha}
	+ \tfrac{\bar b}4 \bar E^{\alpha \dt \alpha}\bar E_{\alpha \dt \alpha}
	+c \bm H^2
	+ (a+c) F^2
	+\Psi^\alpha J_\alpha +\bar \Psi_{\dt \alpha} \bar J^{\dt \alpha}
\cr
L_F &= (a+b) \mathcal W^\alpha \mathcal W_\alpha
	+ (a-b) W^\alpha W_\alpha
\end{align}
in terms of the matter current coupling
\begin{align}
\label{E:abcurrent}
J_\alpha = 2(a+b) \mathcal W_\alpha +2i (a-b)W_\alpha -(a+c) {D}_\alpha F
{~~}.
\end{align}

We now project this action to components. The bosonic Lagrangian is
\begin{align}
\label{E:Component1}
L &=
	\tfrac c2 (\partial_a H)^2
	-\tfrac c2 \tilde H_a^2
	- \tfrac{a+c}2 F_a^2
	- \tfrac{a+c}2 (\partial_a F)^2
	- (a-b) F_{ab}^2
	- (a+b) \mathcal F_{ab}^2
\cr& \quad
	- \tfrac{a}{2} (y_a + \bar y_a)^2
	+ \tfrac{a+c}{2} (y_a - \bar y_a)^2
	+\tfrac b2 {t}^{\alpha \beta} {t}_{\alpha \beta} + \tfrac{\bar b}2  \bar {t}_{\dt \alpha \dt \beta} \bar {t}{}^{\dt \alpha \dt \beta}
\cr& \quad
	+ (y^a + \bar y^a) \hat \jmath_a
	+ i (y^a - \bar y^a) \check \jmath_a
	+\tfrac12 {t}^{\alpha \beta} j_{\alpha \beta} +\tfrac12 \bar {t}_{\dt \alpha \dt \beta} \bar \jmath{}^{\dt \alpha \dt \beta}
~.	
\end{align}
Here we are defining the component currents
\begin{align}
\hat \jmath_a &=  (a+c) F_a + c \,\partial_a H{~}, \qquad
\check \jmath_a = (a+c) \partial_a F  - c \tilde H_a
\cr
j_{\alpha \beta}&= 2 (a+b){\mathcal F}_{\alpha \beta} +2i (a - b) F_{\alpha \beta}
\end{align}
where
\begin{align}
F_{\alpha \beta} &:= {D}_{(\alpha} W_{\beta)}\vert{~}, \qquad
F_{a b} = - (\sigma_{a b})^{\alpha\beta} F_{\alpha\beta} - (\bar\sigma_{a b})^{\dt\alpha\dt\beta} F_{\dt\alpha\dt\beta}{~}, \\
F_{\alpha \dt \alpha} &:= \tfrac12 [{D}_\alpha,\bar {D}_{\dt \alpha}] F\vert{~}, \qquad
\tilde H_{\alpha \dt \alpha} := \tfrac12 [{D}_\alpha , \bar {D}_{\dt \alpha}] H\vert
\end{align}
The auxiliary fields are contained in the second and third lines of
\eqref{E:Component1}, which we denote $L_{aux}$.
Note that if $c=-a$, $\text{Im}\, y_a$ becomes a Lagrange multiplier.\footnote{This is the case in the five-dimensional model \cite{Linch:2002wg}, where $\text{Im}\, y_a$ trivializes the dynamics of the would-be 2-form which is not in the spectrum of five-dimensional fields. This interpretation is confirmed by an analysis of the field strengths of the theory \cite{Gates:2003qi}.
} 
Iintegrating out auxiliary fields (assuming none of $a$, $b$, or $c+a$ vanish), 
\begin{align}
L_{aux} & \to
	- \tfrac{c}{2} (\partial_a H)^2
	+ \tfrac{c^2}{2(a+c)} \tilde H^a \tilde H_a
	+\tfrac{(a+c)^2}{2a} F^aF_a
	+\tfrac{a+c}{2} (\partial_a F)^2
\cr&\quad
	+ \tfrac{m_{11}}2 F_{ab}^2
	+ m_{12} F^{ab}\mathcal F_{ab}
	+ \tfrac{m_{22}}2\mathcal F_{ab}^2
{~~},
\end{align}
where we have dropped total derivative terms like $F\wedge F$.
The coefficients of the gauge field kinetic terms are complicated and given by
\begin{align}
m_{11} &=-\tfrac1{b}(a-b)^2 -\tfrac1{\bar b}(a-\bar b)^2~, \quad
m_{12}=-\tfrac{i}{b}(a^2-b^2) + \tfrac{i}{\bar b}(a^2-\bar b^2) ~,
\cr
m_{22}&=\tfrac1{b}(a+b)^2 +\tfrac1{\bar b}(a+\bar b)^2{~}.
\end{align}
This simplifies when $b^2$ is a real number.
The case of most interest to us (and extended supergravity in general) is when $b$ is a nonzero real number,
\begin{equation}
m_{11}=-\tfrac2{b}(a-b)^2,\quad m_{12}=0,\quad m_{22}=\tfrac2{b}(a+b)^2,\quad\textrm{for}\quad b\in\mathbf{R}^\times~~.
\end{equation}
Including the kinetic terms in the first line of (\ref{E:Component1}) gives
\begin{align}
\label{E:CompRes}
L_{(a,b,c)} =
	-\frac{ac}{2(a+c)} \tilde H_a^2
	+\frac{c(a+c)}{2a} F_a^2
	-\frac ab(a-b) F_{ab}^2
	+\frac ab(a+b) \mathcal F_{ab}^2
{~~}.
\end{align}
The scalar fields $F$ and $H$ have dropped out of the action for any value of the parameters $a$, $b$, and $c$. This is a straightforward consequence of gauge invariance, as in Wess-Zumino gauge, the scalars $F$ and $H$ transform as
\begin{align}
\delta F = -\frac{1}{2} (D^2 \Omega + \bar D^2 \bar\Omega)\vert{~}, \qquad
\delta H = \frac{1}{2i} (D^2 \Omega - \bar D^2 \bar\Omega)\vert~.
\end{align}
They can both be set to zero by a $D^2\Omega\vert$ gauge transformation.
Equivalently, they are always eaten by the auxiliary gauge field $y_a$ (\ref{E:Auxy}), which
is then integrated out.

\subsection{Comments}

The (matter) gravitino multiplet is encoded in a spinor superfield, which is reducible as a representation of the 4D, $N=1$ super-Poincar\'e algebra containing superspins $1\oplus\tfrac12^+\oplus\tfrac12^+\oplus\tfrac12^-\oplus\tfrac12^-\oplus0$ \cite{Gates:1979gv}.\footnote{This language is that of irreducible representations of the four-dimensional super-Poincar\'e algebra. Superspin $1$ contains ordinary spins $(\tfrac32,1)$, $\tfrac12^+$ contains $(1, \tfrac12)$ corresponding to a vector multiplet. Superspins $\tfrac12^-$ and $0$ both correspond to spins $(\tfrac12, 0)$, but the first is a 2-form gauge supermultiplet and the second is a scalar multiplet.
}
(Strictly speaking, retaining all superspins does not give a gauge multiplet.)
At special points in the space of quadratic gravitino Lagrangians, this superspin content is reduced.
When $b=a$, one of the vector multiplets decouples and we recover the results of Butter and Kuzenko.
Further setting $c=-a$ recovers the model of Ogievetsky and Sokatchev, whereas setting $c=0$ gives that of de Wit and van Holten and Fradkin and Vasiliev.
Note that flipping the sign of $b$ simply interchanges the role of the two vector multiplets.
The gravitino sector of the 5D, $N=1$ supergravity model of \cite{Linch:2002wg,Buchbinder:2003qu,Gates:2003qi} has $(a,b,c)=(-1,2,1)$.
Although this latter model was presented with its compensators gauge-fixed, we see from this analysis that these correspond to two vector multiplets and a tensor multiplet with the scalar decoupling since $a+c=0$.
Finally, we found in section \ref{S:Assimilation} that eleven-dimensional supergravity corresponds to the point $(a,b,c)= \tfrac14\times (-1,2,-1)$.
Note that this is a quite generic gravitino multiplet, missing only one superspin $\tfrac12^-$ representation.
These special values are collected in table \ref{F:Gravitini}.

\begin{table}[t]
\begin{align*}
{\renewcommand{\arraystretch}{1.6} 
\begin{array}{|ccrrcc|}
\hline
\textrm{theory}& \textrm{references} & b & c  & \textrm{supergravity} & \textrm{~~~~~superspin \cite{Gates:1979gv}} \\
\hline
\textrm{OS} & $\cite{Ogievetsky:1975vk}$&   a   & -a  & \textrm{4D, }N=2 &  1\oplus\tfrac12^+\oplus\tfrac12^- \\
\textrm{dWvHFV} &$\cite{deWit:1979pq,Fradkin:1979as}$& a& 0 & 	\textrm{4D, }N=2 & 1\oplus\tfrac12^+\oplus0\\
\textrm{LLP} &$\cite{Linch:2002wg,Gates:2003qi}$& -2a& -a &  \textrm{5D, }N=1 & 1\oplus\tfrac12^+\oplus\tfrac12^+\\
\textrm{BBGRL}& \textrm{\cite{Becker:2016edk}} & -2a& a &  \textrm{11D} & 1\oplus\tfrac12^+\oplus\tfrac12^+\oplus\tfrac12^-\oplus0\\
\hline
\end{array}
}
\end{align*}
\caption{Gravitino multiplets}
\footnotesize
The generic spinor superfield contains superspins $1\oplus\tfrac12^+\oplus\tfrac12^+\oplus\tfrac12^-\oplus\tfrac12^-\oplus0$. At special points in the parameter space of gravitino Lagrangians $L_{(a,b,c)}$ (\ref{E:GravitinoLag}) this superspin content is reduced. The two 4D, $N=2$ entries correspond to distinct off-shell 4D, $N=1$ embeddings \cite{Butter:2010sc}.
\label{F:Gravitini}
\end{table}

Returning to the case appropriate to 11D, an important fact is that the ratio $\tfrac ab = -\tfrac12$ is the same for the eleven-dimensional theory as it is for the five-dimensional one. The significance of this is that the Lagrangian depends only on $E_{\un a}^i - \bar E_{\un a}^i$ and not the other linear combination.
We already saw the analogous statement for $B$: The action depends only on the imaginary combination $B-\bar B$ and not the real one so
\begin{align}
L_{(a,-2a,c)} = \int d^4 \theta \, \left[
		-\frac a2 \bm E^{\un a}_i \bm E_{\un a}^i
		+ c \bm H^i \bm H_i
		+(a+c) F_i F^i
		+\Psi^\alpha_i J_\alpha^i
		+\bar \Psi_{\dt \alpha i} \bar J^{\dt \alpha i}
	\right]
\allowdisplaybreaks[0] \cr
	+2a \int d^2 \theta \, \left[
		3 W_i^\alpha  W^i_\alpha
		-  \mathcal W_i^\alpha \mathcal W^i_\alpha
	\right]
\end{align}
where
$\bm E_{\un a i} := E_{\un a i} - \bar E_{\un a i}
	= \bar {D}_{\dt \alpha} \Psi_{\alpha i}
	+ {D}_{\alpha} \bar \Psi_{\dt \alpha i}$.
Both of these statements are important when considering $Y$-dependence of the supergravity gauge parameters $L_\alpha$, because they imply that it is possible to covariantize the conformal supergravity ``mass'' terms $(\partial_i H_a)^2 \to (\partial_i H_a + \bm E_{a i})^2$ by defining the gravitino transformation $\delta \Psi_{\alpha i} \sim 2i \,\partial_i L_\alpha$ \cite{Linch:2002wg}.

{
\footnotesize



\providecommand{\href}[2]{#2}\begingroup\raggedright\endgroup

\end{document}